\documentclass[aps,pra,twocolumn,superscriptaddress,showpacs,floatfix]{revtex4}
\RequirePackage{amsmath}
\RequirePackage{amssymb}
\RequirePackage{xspace}
\RequirePackage{multirow}
\RequirePackage{textcomp}
\newcommand{\dd}{\ensuremath{\text{d}}}
\usepackage{graphicx}

\newcommand{\Eu}{\ensuremath{\text{e}}}
\newcommand{\kb}{\ensuremath{k_\text{B}}}

\newcommand{\fuga}{\ensuremath{\tilde{z}}}

\newcommand{\nex}{\ensuremath{n_\text{ex}}\xspace}
\newcommand{\nc}{\ensuremath{n_\text{c}}\xspace}
\newcommand{\Nc}{\ensuremath{N_\text{c}}\xspace}
\newcommand{\Nex}{\ensuremath{N_\text{ex}}\xspace}

\newcommand{\ie}{\textit{i.\,e.~}}

\newcommand{\cf}{see\xspace}
\newcommand{\insitu}{\textit{in situ}\xspace}

\newcommand*{\microsec}{\ensuremath{\mu s}\xspace}
\newcommand*{\micron}{\ensuremath{\xspace \mu m}\xspace}
\newcommand*{\microk}{\ensuremath{\xspace \mu K}\xspace}
\renewcommand{\Re}[1]{\ensuremath{\text{Re}\left(#1\right)}}
\renewcommand{\Im}[1]{\ensuremath{\text{Im}\left(#1\right)}}
\newcommand{\im}{\ensuremath{i}}
\newcommand{\Tc}{\ensuremath{T_\text{c}}\xspace}
\newcommand{\bra}[1]{\ensuremath{\langle #1|}}
\newcommand{\ket}[1]{\ensuremath{|#1\rangle}}

\newcommand{\Eq}[1]{Eq.~(\ref{eqn:#1})\xspace}
\newcommand{\Eqs}[1]{Eqs.~(\ref{eqn:#1})\xspace}
\newcommand{\Equation}[1]{Equation (\ref{eqn:#1})\xspace}
\newcommand{\Fig}[1]{Fig.~\ref{fig:#1}\xspace}
\newcommand{\Figure}[1]{Figure \ref{fig:#1}\xspace}

\newcommand{\SubSec}[1]{\S~\ref{subsection:#1}\xspace}
\newcommand{\num}[1]{\ensuremath{#1}}
\newcommand{\SI}[2]{\ensuremath{#1}{\rm #2}}
\newcommand{\SIfout}[3]{\ensuremath{\num{#1} \pm \SI{#2}{#3}}}

\newcommand{\numfout}[2]{\num{#1} \ensuremath{\pm} \num{#2}}
\newcommand{\numfoute}[3]{(\num{#1} \ensuremath{\pm} \num{#2})\cdot\num{#3}}

\newcommand{\Na}[1]{\ensuremath{^{#1}}\text{Na}}
\newcommand{\Fg}{\ensuremath{F_\text{g}}}
\newcommand{\Fe}{\ensuremath{F_\text{e}}}
\newcommand{\Mg}{\ensuremath{M_\text{g}}}
\newcommand{\Me}{\ensuremath{M_\text{e}}}

\newcommand{\twoD}{\ensuremath{2}D\xspace}

\newcommand{\mathbold}[1]{\vec{#1}}

\begin{document}
\title{Phase contrast imaging of Bose condensed clouds}
\author{R.~Meppelink, R.~A.~Rozendaal, S.~B.~Koller, J.~M.~Vogels and P.~van der Straten}
\affiliation{Atom Optics and Ultrafast Dynamics, Utrecht University,\\ P.O. Box 80,000, 3508 TA Utrecht, The Netherlands}
\date{\today}

\begin{abstract}
Phase contrast imaging is used to observe Bose-Einstein condensates (BECs) at finite temperature \insitu. The imaging technique is used to accurately derive the absolute phase shift of a probe laser beam due to both the condensate and the thermal cloud. The accuracy of the method is enhanced by using the periodicity of the intensity signal as a function of the accumulated phase. The measured density profiles can be described using a two relevant parameter fit, in which only the chemical potential and the temperature are to be determined. This allows us to directly compare the measured density profiles to different mean-field models in which the interaction between the condensed and thermal atoms is taken into account to various degrees.
\end{abstract}
\maketitle
\section{Introduction}

In the field of ultra-cold atomic physics the two commonly used techniques to image a cloud of atoms are absorption imaging and fluorescence imaging, \ie imaging the cloud's absorbed or radiated intensity, respectively.  In absorption imaging the shadow cast in a probe beam by the cloud of atoms is recorded on a CCD camera. From the resulting image the spatial extent of the cloud and the optical density, proportional to the column density along the line-of-sight of the probe beam, can be derived. For dense clouds absorption imaging becomes unreliable, since for optical densities of the order of $4$ the dynamic range of the absorption imagining technique is insufficient to accurately determine the intensity in the shadowed region \cite{KetterleVarenna1999}. For Bose-Einstein condensates (BECs) the typical \insitu optical density is of the order of \num{500} and the image gets completely ``blacked out'', making it impossible to extract the column density from the image. The fluorescence imaging technique critically depends on the intensity and the frequency of the probe beam. Furthermore, for high densities the cloud becomes opaque and atoms in the center of the cloud are not subjected to the same intensity as atoms on the edge of the cloud. 

One way to reduce the optical density is to reduce the atom-photon cross section by choosing a large probe beam detuning. However, for nonzero detuning the real part of the index of refraction of the atoms becomes nonzero and the cloud behaves like a gradient-index lens. This results in a refraction of the probe beam and leads to distortion of the images. The resulting images cannot be used for a relative or absolute column density measurement, since the magnitude of this effect has a spatial dependence. Detuning the probe beam further from resonance reduces the index of refraction but leaves the cloud virtually transparent before the refraction is sufficiently reduced. 

The optical density can successfully be decreased by reducing the atomic density. The cloud is allowed to expand during a specific time-of-flight (tof) and an image is recorded as soon as the optical density is of the order of \num{2}. Although absorption imaging after expansion is the commonly used technique in this field, expansion of the cloud introduces some difficulties as well. First, if the cloud is a BEC at finite temperature there is no accurate description available of the expansion of the cloud. Second, the quantization axis of the cloud is no longer well-defined since the magnetic fields are turned off. As a result, the effective cross section for the polarized probe beam becomes ambiguous. As a consequence both absorption imaging and fluorescence imaging cannot be applied to accurately determine the column density of dense clouds.

Some techniques have been developed to image dense clouds by using strong saturation in both absorption \cite{OptLett.32.3143} and fluorescence \cite{OptComm.180.73} imaging. However, the high-intensity regime has to be calibrated based on the low-intensity regime and the latter regime is imaged after expansion. As a consequence the issues raised above about imaging after expansion apply to the strong saturation techniques as well, although cloud can be imaged \insitu. 

Alternatively, phase contrast imaging (PCI) has been successfully applied to image dense atomic clouds \cite{PhysRevLett.78.985,Science.273.84,PhysRevLett.79.553}. In PCI the real part of the index of refraction is measured. The detuning can be chosen in such a way that the refraction is small but the phase shift due to the index of refraction remains substantial. PCI can be used to image BECs in a nondestructive way, since the number of scattered photons becomes negligible for large detuning. Other techniques based on dispersive light scattering, such as diffraction-contrast imaging of cold atoms \cite{PhysRevA.72.031403} and nondestructive spatial heterodyne imaging of cold atoms \cite{OptLett.26.137} have been demonstrated on dense thermal clouds.

Here, we use PCI to measure the \insitu density distribution of dense BECs at finite temperature absolutely and accurately. By choosing a smaller detuning compared to the nondestructive PCI schemes, the density of the thermal cloud and the BEC are determined accurately for all typical temperatures. This allows us to directly study the interplay between the condensed and thermal atoms and the influence of this interaction on the transition temperature for condensation.

This paper begins with a detailed description of the PCI technique and the way it is applied to a cold Bose gas. Next, the accuracy is demonstrated  in two experiments. In the first experiment, described in \SubSec{problemsabsim}, the number of condensed atoms is measured with an accuracy below \SI{5}{\%} in a single shot. In the second experiment, described in \SubSec{interactions} the growth of the condensate is studied as a function of the temperature. In this measurement, the interaction between the condensed and thermal atoms is studied directly. Furthermore, measurements around the temperature where a BEC is formed are used to determine the transition temperature.

\section{Theoretical description of PCI}
\subsection{Imaging the phase}
\label{subsection:imagingthephase}
It is possible to image a cloud of atoms by illuminating the cloud with a probe beam and record the shadow cast due to the absorption of the atoms on a CCD camera. Since the optical density is measured, this technique probes the imaginary part of the complex index of refraction $\mathcal{N}$ of the atomic cloud. In contrast, the imaging technique described in this paper probes the real part of $\mathcal{N}$.  This technique is used in the phase contrast microscope which is developed by Zernike in 1933. In 1953 Zernike was awarded the Nobel prize ``for his demonstration of the phase contrast method, especially for his invention of the phase contrast microscope'' \cite{Zernike}.

In general, the complex phase $\phi$ accumulated by the light in a medium such as a cloud of atoms can be expressed as $\phi_\text{atoms}=\phi^{'} + \im \phi^{''}/2$, with $\phi^{''}$ the optical density. Writing the electric field of the probe beam as $E_\text{probe}=E_0 \exp(\im \phi)$ with $E_0$ the amplitude of the light field, the electric field after passing a cloud of atoms can be written as
\begin{equation}
E=E_\text{probe}+E_\text{atoms}=E_\text{probe} + E_\text{probe}(\Eu^{\im \phi_\text{atoms}}-1),
  \label{eqn:phase}
\end{equation}
where the electric field is split in two parts: the part that is diffracted by the atoms, $E_\text{atoms}$, and the part that is not, $E_\text{probe}$. In absorption imaging the phase information is lost on a CCD camera, since the intensity $I$ is measured as $I=c \varepsilon_0 EE^*/2 = I_0e^{-\phi^{''}}$ where $I_0 = c \varepsilon_0 |E_0|^2/2$, $c$ is the speed of light and $\varepsilon_0$ is the vacuum permittivity. So the real phase term $\phi^{'}$ cancels and only the absorption is measured. In the simplest form of PCI, the non-diffracted  part of the probe light (first term on the RHS of \Eq{phase}) is blocked. The light that arrives at the camera is the light which is diffracted by the atoms, $E=E_\text{probe} (\exp\left ( \im \phi_\text{atoms} \right )-1)$. Now the phase information no longer cancels and the intensity on the camera is given by
\begin{equation}
  I=2 I_0 (1-\cos{(\phi_\text{atoms})}),
  \label{eqn:darkfield}
\end{equation}
where the arbitrary phase of the probe beam cancels. This technique is referred to as dark-field imaging and has been successfully used to image \Na{23} atoms in the early days of experimental BEC physics \cite{PhysRevLett.79.553}. An important experimental drawback of this method is the absence of probe light on the camera preventing the normalization of the intensity profile, which makes this technique not suitable to measure the absolute phase shift. However, the relative signal can be used to study the spatial distribution.

A more elaborate way of converting the accumulated phase into an intensity profile is by phase shifting the non-diffracted light instead of blocking it, in the same way as the phase contrast microscope works. The non-diffracted light is the plane wave part of the beam, which can be phase-shifted by placing a small transparent object with a refractive index in the probe beam in such a way that only the plane wave part of the beam propagates through it.  This object will be called a phase spot. The plane wave part propagating through the phase spot accumulates a phase $\theta$, changing \Eq{phase} to
\begin{equation}
  \label{eqn:phase_phasespot}
E=E_\text{probe} \Eu^{\im \theta} + E_\text{probe}(\Eu^{\im \phi_\text{atoms}}-1). 
\end{equation}
Now, the intensity is given by
\begin{equation}
I = I_0 (3-2 \cos(\theta)+2 \cos(\theta-\phi_\text{atoms})-2 \cos(\phi_\text{atoms})),
  \label{eqn:phase_phasespot_full}
\end{equation}
where the arbitrary phase of the probe beam cancels. The probe beam is not blocked and it is possible to derive the absolute phase shift, since the intensity profile $I_0$ can be measured and used for normalization.  This technique has also been successfully applied to image \Na{23} atoms \cite{Science.273.84}. The implementation of PCI described in Ref.~\cite{Science.273.84} uses a detuning of the probe beam, which is large compared to the atomic resonance and the absorption of the light by the atoms is therefore reduced to a degree that it is no longer destructive. This allows for taking multiple images of the same BEC. 

In our experiment we do not focus on the ability to make a nondestructive image but we use the periodicity of the intensity as a function of $\phi_\text{atoms}$ (\cf \Eq{phase_phasespot_full}) to accurately determine the accumulated phase of both the BEC and the thermal cloud. In recent years, different schemes are proposed derive the phase information of a cloud of cold atoms \cite{Turner2004,OptLett.26.137}, but the accumulated phase has never been used for accurate quantitative measurements of the density distribution of BECs at finite temperature.  

\subsection{Index of refraction of a cold gas}
We continue with a detailed description of the phase shift of an electro-magnetic wave propagating through a cloud of spin-polarized atoms in order to relate $\phi_\text{atoms}$ to the atomic density.  The phase shift is caused by the modification of the probe field by the atomic dipoles, so we start with an expression of the polarizability of one atom. The polarizability tensor $\overleftrightarrow{\alpha}$ in the low-intensity limit and $\delta \ll \omega$ is given by \cite{Nienhuis1991}

\begin{equation}
\label{eqn:nienhuis}
\overleftrightarrow{\alpha} = \sum_{g,g',e}\frac{\im}{\hbar} \frac{1}{\gamma/2-\im \delta}\bra{g}\vec{\mu}_{\text{ge}}\ket{e}\bra{e}\vec{\mu}_{\text{eg}}\ket{g'}\bra{g'}\sigma_{\text{gg}}\ket{g},
\end{equation}
where the natural linewidth $\gamma$ is given by
\begin{equation}
\gamma = \frac{\omega^3 \mu^2}{3 \pi\epsilon_0 \hbar c^3}.
\label{eqn:def-gamma}
\end{equation} 
Here, $\vec{\mu}_{{eg,ge}}$ are the electric-dipole operators, $\sigma_\text{gg}$ is the density matrix, $\delta$ is the detuning and $\omega$ is the probe frequency.
In the experiment various linear polarizations of the light field have to be dealt with, so the dependence of the polarizability $\overleftrightarrow{\alpha}$ on the polarization has to be determined. We define the angle $\beta$ as the angle between the $z$-axis of the atomic cloud and the (linear) polarization direction of the light field which is in the $(x,z)$ plane. In this system, the line-of-sight is along the $y$-axis. In order to evaluate \Eq{nienhuis} for different polarizations, a fixed polarization axis is chosen parallel to the polarization of the light field and \Eq{nienhuis} is evaluated in this frame, in which conveniently only $\Delta M=0$ transitions are induced. Furthermore, only one component of the polarizability tensor will contribute in this frame and the polarizability becomes a scalar. Since all atoms in the trap are in the $\ket{\Fg,\Mg}=\ket{1,-1}$ state with respect to the $z$-axis of the magnetic field confining the atoms, the density matrix $\sigma_{gg}$ only has one nonzero element, which makes the calculation of the rotation from the magnetic $z$-axis to the polarization axis of the light field straight-forward.

Rotating the density matrix of the ground state $\sigma_{gg}$ is achieved using
\begin{equation}
\label{eqn:dens_mat_rot}
\hat{\sigma}_{gg}(\beta)=\mathcal{R}_y(\beta)^\dagger \sigma_{gg} \mathcal{R}_y(\beta),
\end{equation}
 where the rotation matrix $\mathcal{R}_y(\beta)$ for rotating a system around the $y$-axis over an angle $\beta$ for a system with $J_g=1$ can be found using Wigner's formula as \cite{Sakurai}
\begin{equation}
 \mathcal{R}_y(\beta)=
 \left( \begin{array}{ccc}
\cos^2(\frac{\beta}{2}) & \frac{\sin(\beta)}{\sqrt{2}} & \sin^2(\frac{\beta}{2})  \\
-\frac{\sin(\beta)}{\sqrt{2}}  & \cos(\beta) & \frac{\sin(\beta)}{\sqrt{2}}  \\
 \sin^2(\frac{\beta}{2})  & -\frac{\sin(\beta)}{\sqrt{2}}    & \cos^2(\frac{\beta}{2})  \end{array} \right) ,
\end{equation}

Since the two dipole moment operators $\vec{\mu}_{eg}$ do not change the $M$-state of the atoms, only the diagonal elements of \Eq{nienhuis} play a role and we end up with 
\begin{equation}
  \alpha = \frac{2 \im}{\hbar \gamma} \sum_{g,e}{\frac{\bra{\Fg,\Mg}\vec{\mu}_{ge}\ket{\Fe,\Me}^2  \bra{\Fg,\Mg}\hat{\sigma}_{gg}(\beta)\ket{\Fg,\Mg}}{1-2 \im \delta_{{e}}/\gamma}}.
\end{equation}
\noindent Furthermore, we can define the square of the Clebsch-Gordan coefficients $C_{\Fe,\Mg}$ (transition strength) as
\begin{equation}
  C_{{g},{e}} = \bra{\Fg,\Mg}\vec{\mu}_{{ge}}\ket{\Fe,\Me}^2 / \mu^2,
\end{equation}
and obtain using \Eq{def-gamma}
\begin{equation}
  \alpha = \frac{\im \epsilon_0 c \sigma_{\lambda}}{\omega} \sum_{g,e}{\frac{C_{{g},{e}} \bra{\Fg,\Mg}\hat{\sigma}_{{gg}}(\beta)\ket{\Fg,\Mg}}{1-2 \im \delta_{{e}}/\gamma}},
\end{equation}
\noindent where $\delta_{e}$ is the detuning with respect to transition $\Fg \rightarrow \Fe$ and $\sigma_{\lambda}$ the cross section for absorption of light:
\begin{equation}
\sigma_{\lambda} \equiv \frac{3 \lambda^2}{2 \pi}.
\end{equation}

For sodium atoms in the $\ket{\Fg,\Mg}=\ket{1,-1}$ state this results in 
\begin{equation}
\label{eqn:polphi}
\alpha =\frac{\im \epsilon_0 c \sigma_{\lambda}}{\omega}  \sum_{e}{\frac{\mathcal{D}_{\Fe}(\beta)}{1-2 \im \delta_{e}/\gamma}},
\end{equation}
where 
\begin{subequations}
\begin{align}
\mathcal{D}_{0}(\beta) &= \frac{4}{24} \sin^2(\beta), \\
\mathcal{D}_{1}(\beta) &= \frac{5}{24}(1+\cos^2(\beta))+\frac{6}{24}, \\
\mathcal{D}_{2}(\beta) &= \frac{6}{24}+\frac{1}{24}\sin^2(\beta). 
  \label{eqn:d0d1d2}
\end{align}
\end{subequations}

For alkali metal atoms the polarizability is independent of the angle $\beta$ if the detuning $\delta$ is large compared to the hyperfine splitting of the excited state. In that limit the numerator of the sum in \Eq{polphi} is $\frac{2}{3}$ with no angular dependence.
Finally, the complex index of refraction $\mathcal{N}$ is given by 
\begin{equation}
\label{eqn:complexior}
\mathcal{N}^2=1+\frac{\rho \alpha}{\epsilon_0},
\end{equation}
with $\rho$ the density. Note that for atoms confined in a potential the density is not homogeneous and $\rho$ depends on the position, $\rho=\rho(x,y,z)$.

The index of refraction given by \Eq{complexior} is only valid for low densities. For higher densities, the dipole moment of an atoms is influenced by the internal field of the surrounding atoms. This results in a modification of the refractive index, which is accounted for by the Lorentz-Lorenz equation \cite{Jackson}   
\begin{equation}
  \mathcal{N}^2 = 1+\frac{\rho \alpha/\varepsilon_0}{1+C},
\label{eqn:modiclas}
\end{equation}
where $C$ is given by
\begin{equation}
  C = -\frac{1}{3} \rho \alpha/\varepsilon_0. 
  \label{eqn:collective_class}
\end{equation}

Since the atoms being imaged are bosons instead of classical particles, the index of refraction will be modified further due to bunching of the atoms. Two terms are added to \Eq{collective_class}. The first term accounts for modifications due to the enhanced photon scattering cross section caused by the quantum statistics of the atoms. The second term describes the modified refractive index induced by resonant Van der Waals interaction, resulting in an increase of almost $10\%$ in both the real part and the imaginary part of the refractive index \cite{Morice1995}.

We have derived the modifications along the lines of Ref.~\cite{Morice1995} and have confirmed the result for the detuning and densities reported in that paper. We found modifications to the complex index of refraction strongly depend on the chosen detuning and therefore on $\alpha$. In Ref.~\cite{Morice1995}, $\delta \sim \gamma$, while we are in the regime $\delta \gg \gamma$. For our parameters, the real part of the index of refraction does not change significantly. However, we do find an enhancement of the absorption by a factor of $3$. Since the overall absorption is very small, up to four percent at the detuning used, enhanced absorption does not influences the intensity profile significantly. As a result, we expect the measured phase to change less than one percent due to this effect.

\subsection{Imaging the atomic density}

The image of a cloud of atoms recorded on a CCD camera yields the column density of the cloud, that is the density profile integrated along the probe line-of-sight. The column density can be linked to the complex index of refraction in the following way.
In general, light passing through a medium with an index of refraction $\mathcal{N}$ over a distance $l$ accumulates a complex phase $\phi$  given by 
\begin{equation}
  \phi = k(\mathcal{N}-1)\,l,
\end{equation}
\noindent where $k=2 \pi/\lambda$. Since the index of refraction depends on the position, the accumulated phase is written as
\begin{equation} 
\label{eqn:phi_cor}
\phi(x,z) = k\int  \left ( \sqrt{1+ \frac{\rho(x,y,z) \alpha /\varepsilon_0}{1+C}}-1 \right ) \dd y. 
\end{equation} 
Since $\left | \rho \alpha/\varepsilon_0 \right | \ll 1$ the index of refraction can be approximated as
\begin{equation}
  \mathcal{N}=\sqrt{1+ \frac{\rho \alpha/\varepsilon_0}{1+C}} \approx 1+\frac{\rho \alpha}{2 \varepsilon_0}.
\end{equation}
Under typical experimental conditions $\left | \rho \alpha/\varepsilon_0 \right |$ is $10^{-2}$, introducing only a small error by using this approximation. Now, \Eq{phi_cor} can be written as
\begin{equation}
\label{eqn:realphase}
\phi(x,z)= k \, \frac{\alpha}{2 \varepsilon_0}\int \rho(x,y,z) \dd y =  k \, \frac{\alpha}{2 \varepsilon_0} \rho_{\text{c}}(x,z),
\end{equation}
\noindent where the integration is along the line-of-sight $y$ and $\rho_c(x,z)$ is the column density. 

The real part of $\phi(x,z)$, proportional to $\Re{\alpha}$, yields the phase shift, whereas the imaginary part, proportional to $\Im{\alpha}$, yields the absorption. 
Since $\alpha$ is constant for fixed detuning $\delta$, the phase shift is directly proportional to the column density. Since \Eq{polphi} satisfies the Kramers-Kronig relation 
\begin{equation}
\Re{\mathcal{N}} -1 = \left(\frac{2 \delta}{\gamma}\right) \Im{\mathcal{N}},
\end{equation} 
the ratio between the phase shift and the absorption is constant for a given detuning.
\section{PCI setup}
\begin{figure}
  \begin{center}
    \includegraphics[width=0.9\columnwidth]{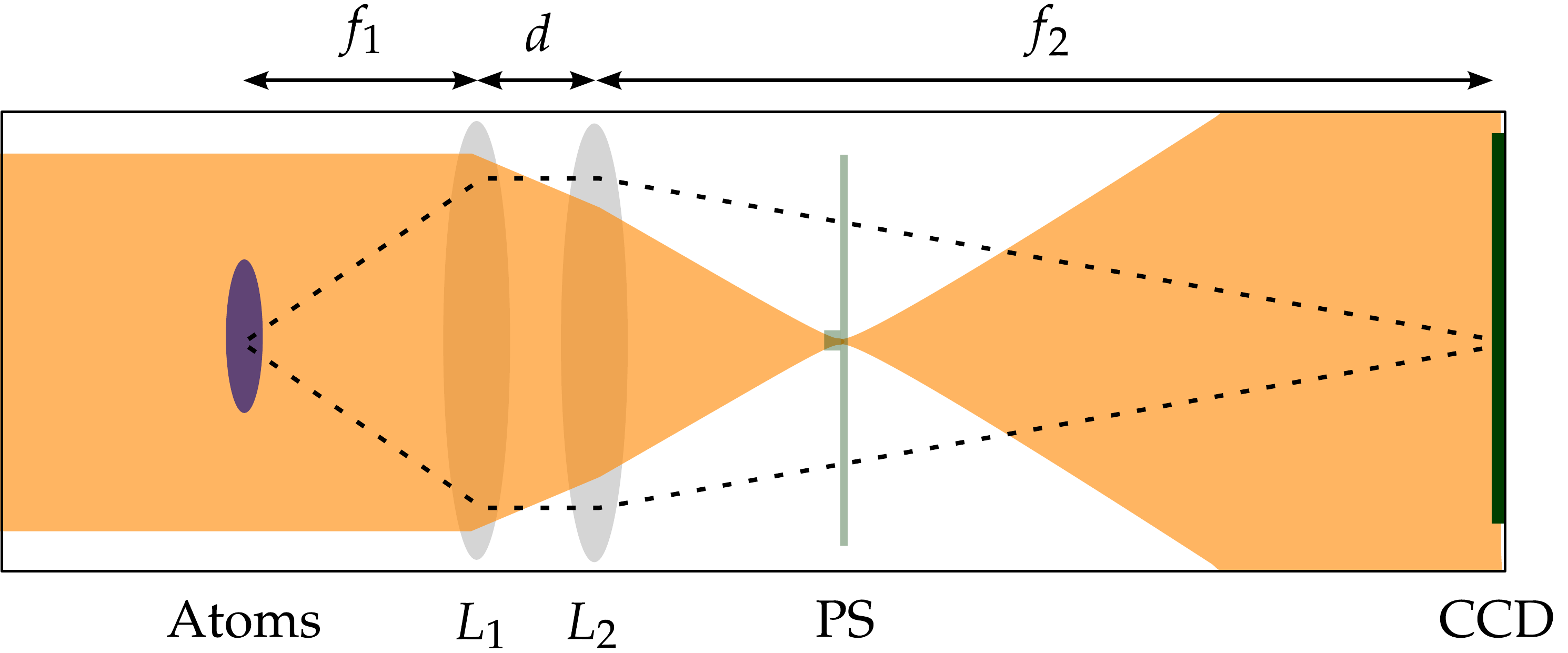}
  \end{center}
  \caption{Schematic representation of the PCI setup. The atoms located at the center of the trap are imaged with lens $L_1$ with focal distance $f_1$ which is placed at a distance $f_1$ from the center. A sharp image is created by placing the lens $L_2$ at distance $d$  from $L_1$ with focal length $f_2$ at a distance $f_2$ from the CCD camera. The phase spot (PS) is placed in the focal plain of the non-diffracted probe beam.}
  \label{fig:pci-setup}
\end{figure}

In order to create a phase contrast image of the atoms, the plane wave part of the probe field has to be phase shifted as is described in \SubSec{imagingthephase}. This is commonly achieved by Fourier transformation using a lens. The concept of the imaging setup used in our experiment is schematically shown in \Fig{pci-setup}. A two-lens setup is used to generate a sharp image of the atoms on a camera. The first lens $L_1$ with focal distance $f_1=\SI{250}{mm}$ is placed at a distance $f_1$ from the center of the magnetic trap. The second lens $L_2$ with focal distance $f_2=\SI{750}{mm}$ is placed at a distance $f_2$ from the CCD camera. The distance between both lenses is $d=\SI{150}{mm}$. The combination of both lenses creates a sharp image of the cloud on the CCD camera with a magnification $M=f_2/f_1=\numfout{3.0}{0.05}$ where the uncertainty is the result of the uncertainty in the focal distance of the lenses.

In order to shift the phase of the plane wave part of the light a phase spot is placed in the back focal plane (Fourier plane) of the two lenses, which lies for these parameters beyond the second lens $L_2$. This position can be found easily in the experiment, since the parallel probe beam has its waist at the back focal plane and therefore the phase spot is placed at this position.

\begin{figure}
  \begin{center}
    \includegraphics[width=0.75\columnwidth]{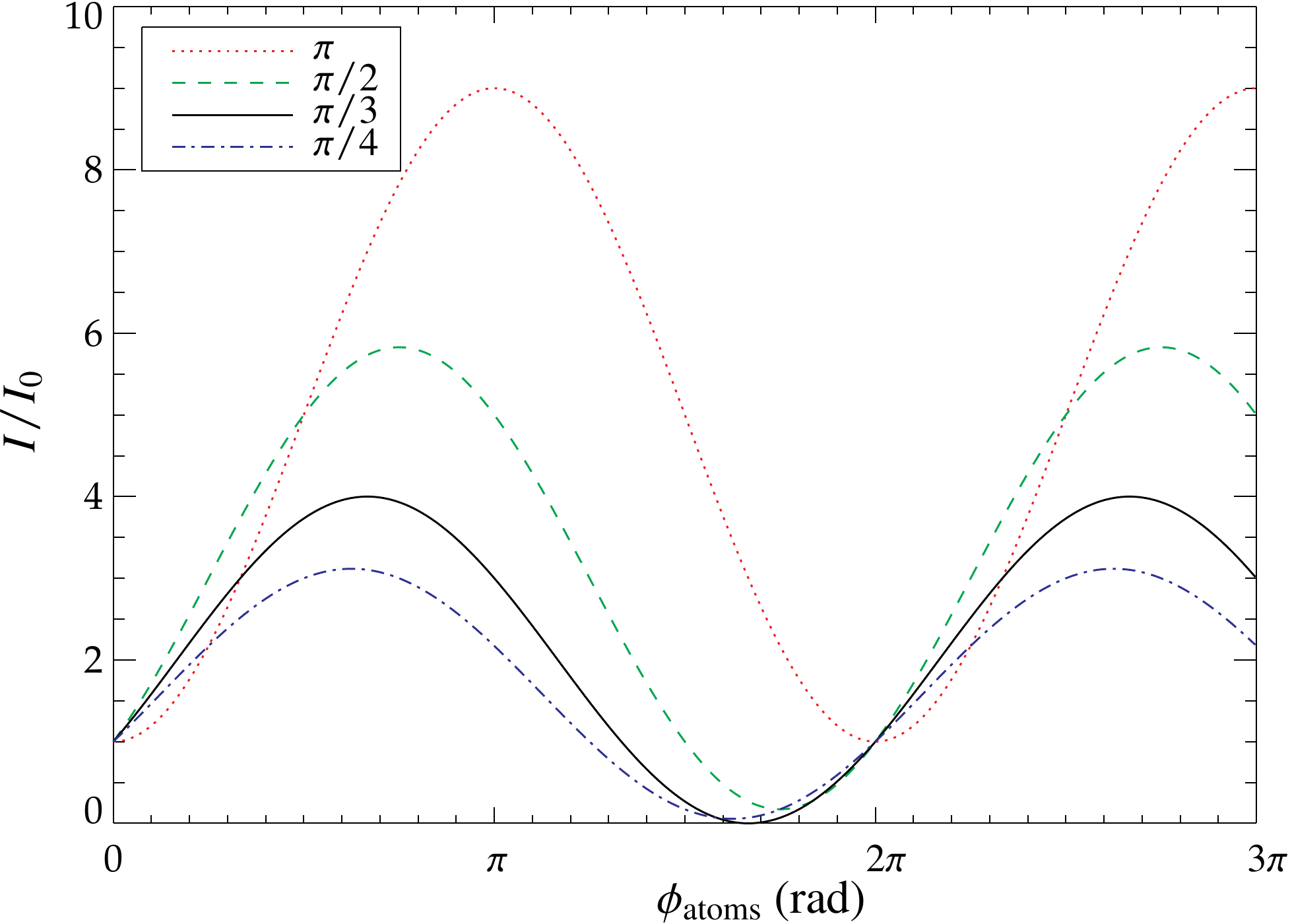}
  \end{center}
  \caption{The phase contrast signal $I/I_0$ as a function of the accumulated phase $\phi_\text{atoms}$ for different phases $\theta=\pi,\pi/2,\pi/3,\pi/4$ of the phase spot.}
  \label{fig:phase-phi-theta}
\end{figure}

\Figure{phase-phi-theta} shows the intensity $I/I_0$ as a function of the accumulated phase $\phi_\text{atoms}$ for different values of $\theta$ based on \Eq{phase_phasespot_full}. In all experiments described here, $\theta=\pi/3$ is used. This value is chosen since it yields a maximum visibility $(I_{\text{max}}-I_{\text{min}})/(I_{\text{max}}+{I_{\text{min}}})=1$ and a dynamic range $ I_{\text{max}} - I_{\text{min}} = 4I_0$. A visibility of \num{1} is reached only for $\theta=\pi/3$, as can be seen in \Fig{phase-phi-theta}. Furthermore the response is approximately linear for small accumulated phases. 

For technical reasons, the $\pi/3$ spot consists of a glass plate with a dimple instead of a spot on top. This dimple is made by dry-etching a \SI{50}{\micron} diameter round dimple with a depth of $5 \lambda/(6(\mathcal{N}-1))=\SIfout{1071}{15}{nm}$ in a \SI{50}{mm} round fused silica glass plate of \SI{4}{mm} thickness \cite{thxmlf}. Light propagating through the dimple accumulates $5\pi/3$ less phase than the light propagating through the full thickness of the plate and thus $\theta=+\pi/3$.

The atoms are imaged using an Apogee AP1E camera with a Kodak KAF-0401E chip camera with a pixel size of $9\times9\,{\micron}^2$. At the magnification $M=3.0$, the effective camera resolution of $3.0\,\micron$ per pixel is comparable to the diffraction limit $x_\text{res} \approx 1.22 \lambda f_1 /(2 r)\approx 3.6 \, \micron$ of the imaging lens $L_1$ with radius $r=\SI{25}{mm}$ for probe light with a wavelength $\lambda = \SI{589}{nm}$.

\section{Density distribution}
\subsection{Three models for BECs at finite temperature}
\label{subsection:threemodels}
A phase contrast image of a BEC at nonzero temperature contains the column density distribution of the two-component cloud, \ie a BEC and a thermal cloud which can be distinguished from each other due to their distinct density distributions. In order to derive the relevant parameters from such an image, like the thermal (excited) density \nex, the condensate (ground state) density \nc, and the temperature $T$, the measured profiles are fitted to a theoretical model describing the cloud. Various models exist in literature in which the interaction between the condensed and thermal atoms are taken into account to various degrees. In this paper we compare our measurements to three of these models. The first model is the commonly used bimodal distribution in which all interactions between the two components are ignored. This model is used in to describe the vast majority of the experiments of cold Bose gases \cite{KetterleVarenna1999}.  In the second model only the effect of the condensed atoms on the thermal atoms is taken into account. The third model incorporates the effect of the density distribution of the thermal atoms on the condensed atoms as well. Below, the density distribution of the thermal cloud and the BEC in each of these models are described consecutively. 

In the first model, referred to as the ideal model, the chemical potential is given by $\mu= V_\text{ext}(\mathbold{r})+\nc(\mathbold{r})U_0$, where $U_0=4 \pi \hbar^2 a/m$ is the interaction parameter for bosons with $s$-wave scattering length $a$ and mass $m$ \cite{BECIDG}.  Since there are no interactions between condensed atoms with density $\nc$ and thermal atoms with density $\nex$, $\mu$ does not depend on $\nex$. In the ideal model $\nex$ is given by 
\begin{equation}
  \nex(\mathbold{r}) = \int \frac{\dd \mathbold{p}}{(2 \pi \hbar)^3} \frac{1}{\Eu^{(\varepsilon(\mathbold{r})-\mu)/\kb T}-1},
  \label{eqn:nexBog}
\end{equation}
where $\varepsilon$ is given by
\begin{equation}
  \varepsilon(\mathbold{r}) = \frac{p^2}{2m} +V_\text{ext}(\mathbold{r}).
  \label{eqn:epspBogol}
\end{equation}
In this approximation both $\nex$ and $\nc$ can be derived analytically yielding the bimodal density distribution, the sum of a Maxwell-Bose distribution modeling the thermal cloud and a Thomas-Fermi (TF) distribution modeling the BEC. In this approximation $\nex$ is given by 
\begin{equation}
  \nex(\mathbold{r}) =  \text{Li}_{3/2}\left(\fuga\Eu^{-V_\text{ext}(\mathbold{r})/\kb T}\right)/\lambda^{3}_\text{dB},
  \label{eqn:neqBogExpr}
\end{equation}
where the de Broglie wavelength is given by $\lambda_\text{dB}=\sqrt{(2 \pi \hbar^2)/(m \kb T)}$ and $\fuga=\exp(\mu/(\kb T))$ is the fugacity. $\text{Li}_{3/2}(z)$ is the Bose function, given by  $\text{Li}_{n}(z)\equiv \sum_{k=1}^\infty z^k/k^n$ and 
\begin{equation}
  V_\text{ext}=\frac{1}{2} m (\omega^2_\text{rad} r_x^2 + \omega^2_\text{rad} r_y^2 +\omega^2_\text{ax} r_z^2),
  \label{eqn:Vext}
\end{equation}
is the external potential. In the TF approximation $\nc$ is given by
\begin{equation}
  \nc(\mathbold{r})=\frac{\mu}{U_0}\left[ 1-\left(\frac{r_x}{R_\text{rad}}\right)^2 - \left(\frac{r_y}{R_\text{rad}}\right)^2- \left(\frac{r_z}{R_\text{ax}}\right)^2  \right],
  \label{eqn:n0}
\end{equation}
where $R_\text{rad,ax}$ are the TF radii of the condensate given by
\begin{subequations}
\begin{align}
  R_\text{rad}&=\sqrt{\frac{2\mu}{m\omega^2_\text{rad}}},\\
  R_\text{ax}&=\sqrt{\frac{2\mu}{m\omega^2_\text{ax}}}.
  \label{eqn:TFrmu}
\end{align}
\end{subequations}
Since $\omega_\text{rad}/\omega_\text{ax} \approx 50$ in the experiments described here and thus  $R_\text{rad} \ll R_\text{ax}$ the clouds are cigar shaped. The number of condensed atoms $\Nc$ is found by integrating $\nc$ over $\mathbold{r}$ and yields
\begin{equation}
  \Nc=\int \nc(\mathbold{r}) \dd \mathbold{r} = \frac{8 \pi}{15} \left( \frac{2 \mu}{m \bar{\omega}^2} \right)^{3/2} \frac{\mu}{U_0}.
  \label{eqn:nbec}
\end{equation}
Integration of the density over the line-of-sight yields the column density, which is the property being measured in the experiments. 

The second model, denoted as the semi-ideal model, does include interactions, but only the contribution of the mean-field potential of the condensate to the thermal density distribution is taken into account \cite{PhysRevA.58.2423}. Therefore, $\mu$ remains unchanged: $\mu=V_\text{ext}(\mathbold{r})+\nc(\mathbold{r}) U_0$, but the energy of the thermal atoms is changed due to the mean-field potential of the condensed atoms and written as
\begin{equation}
\varepsilon^2(\mathbold{r}) =  \left( p^2/(2 m) +2 \nc(\mathbold{r}) U_0 +V_\text{ext}(\mathbold{r})-\mu \right)^2-\left( \nc(\mathbold{r})U_0 \right)^2.
  \label{eqn:energySI}
\end{equation}

\begin{figure}
  \begin{center}
    \includegraphics[width=0.65\columnwidth]{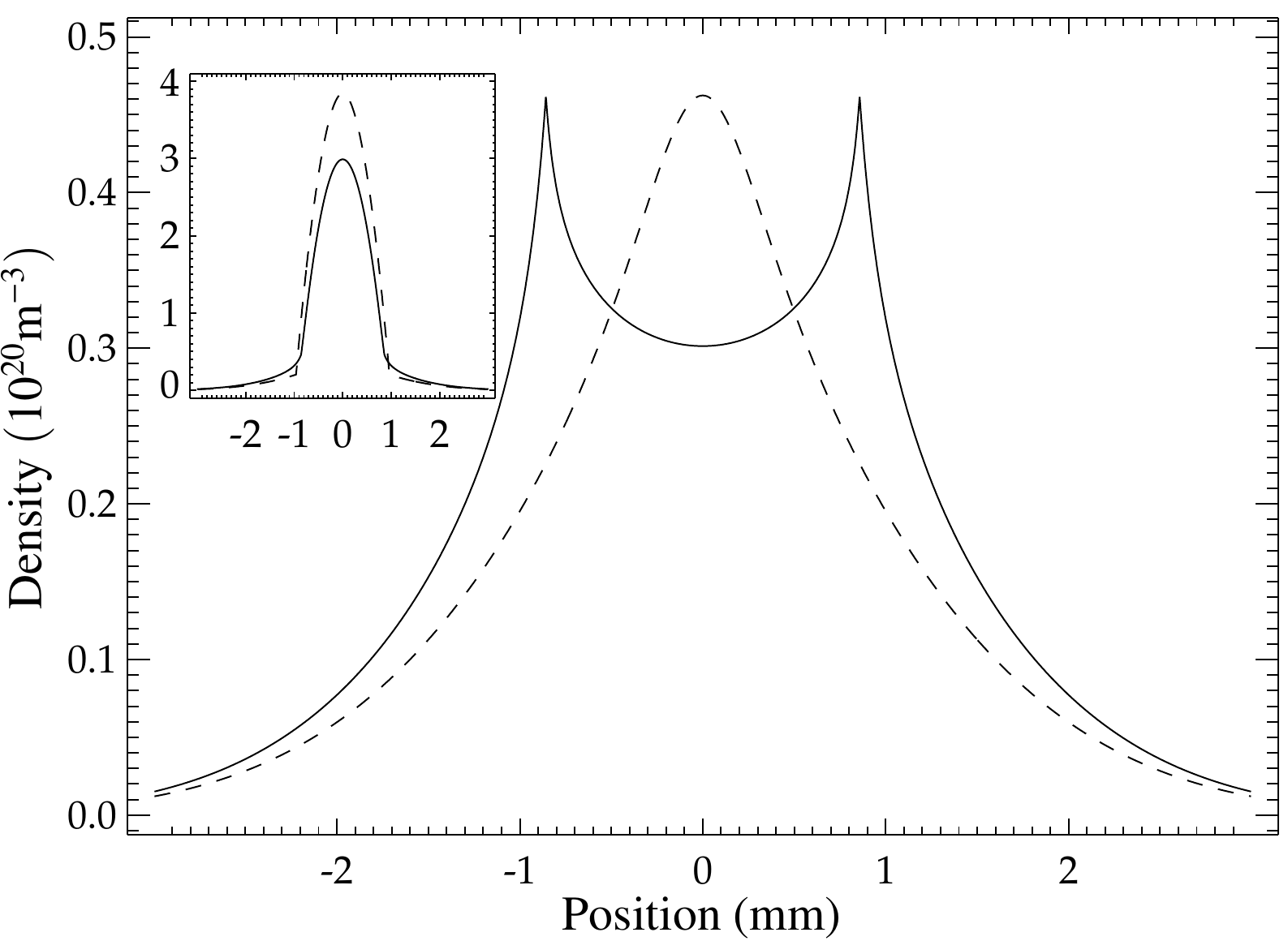}
  \end{center}
    \caption{The density distribution of a thermal cloud $\nex$ calculated using the noninteracting model (dashed line) and the semi-ideal model (solid line), for $T=\SI{0.9}{\microk}$, $N=\SI{1\cdot10^9}{atoms}$ for the typical experimental trap parameters. The inset shows the total density distribution $\nex+\nc$ for both models.}
   \label{fig:bimodal}
\end{figure}

We calculate the density distribution in this semi-ideal model by numerical integration of \Eq{nexBog}, where $\varepsilon(\mathbold{r})$ is given by \Eq{energySI} \cite{BECIDG}. The shape of the density distribution of the condensate remains unchanged, but the repulsion of the mean-field potential of the condensate causes the thermal atoms to be repelled from the center, which results in the shape shown in \Fig{bimodal}. In order to have the same total number of atoms and temperature in \Fig{bimodal}, the chemical potential is different in both models. The decrease in density in the center is the result of the repulsion due to the mean-field potential of the condensed atoms and therefore absent in the noninteracting distribution. Due to the lower central density the resulting condensate fraction is lower in the semi-ideal model compared to the noninteracting model for the same total number of atoms $N=\Nex+\Nc$ and temperature $T$.

\begin{figure}
  \begin{center}
    \includegraphics[width=0.65\columnwidth]{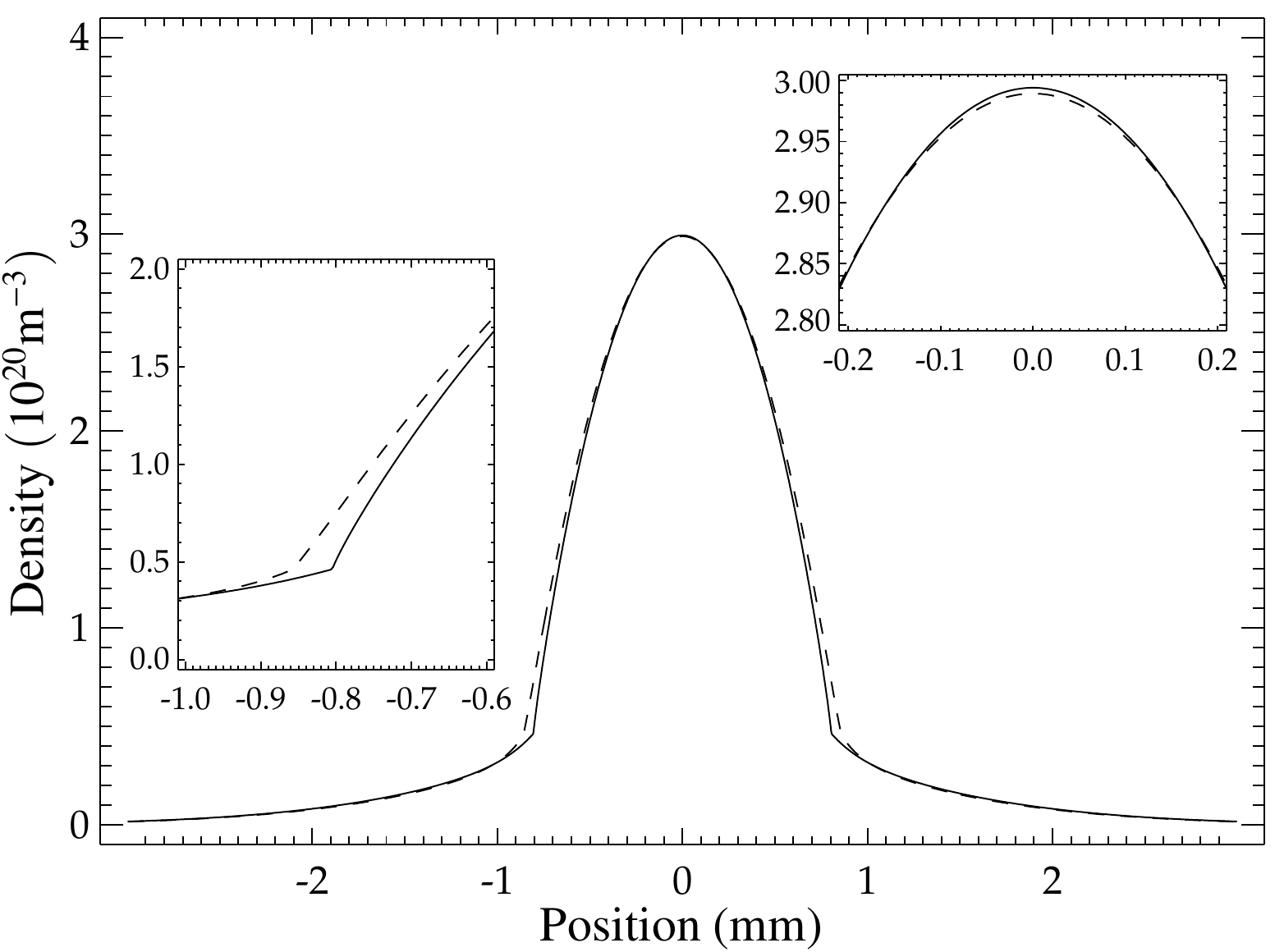}
  \end{center}
    \caption{The combined density distribution of the thermal cloud and condensate $\nex + \nc$ calculated using the semi-ideal model (dashed line) and the Popov model (solid line) for $T=\SI{0.9}{\microk}$, $N=\SI{1\cdot10^9}{atoms}$ and the typical experimental trap parameters. The left and right insets show an enlargement of the edges and the center of the condensate, respectively.}
  \label{fig:bimodalBenP}
\end{figure}

In the third model the mean-field effect of the thermal cloud on the condensate is taken into account as well. This model corresponds to the Popov approximation in the TF limit, so we refer to this model as the Popov model \cite{BECIDG}. The chemical potential is now given by $\mu=V_\text{ext}(\mathbold{r})+\left[ \nc(\mathbold{r})+2\nex(\mathbold{r}) \right]U_0$. The expression for the density of the excited atoms in the Popov approximation is given by
\begin{equation}
  \nex(\mathbold{r}) = \int \frac{\dd \mathbold{p}}{(2 \pi \hbar)^3}\frac{p^2/2m +2 n(\mathbold{r}) U_0 +V_\text{ext}(\mathbold{r})-\mu}{\varepsilon(\mathbold{r})}\frac{1}{\Eu^{\varepsilon(\mathbold{r})/\kb T}-1},
  \label{eqn:nexPopov}
\end{equation}
with $n=\nex+\nc$ and where, in the semi-classical approximation, $\varepsilon(\mathbold{r})$ is given by
\begin{equation}
  \varepsilon^2(\mathbold{r}) = \left( p^2/(2m) +2 n(\mathbold{r}) U_0+V_\text{ext}(\mathbold{r})-\mu \right)^2-\left( \nc(\mathbold{r})U_0 \right)^2.
  \label{eqn:epspPopov}
\end{equation}
Now, the resulting density distribution of both $\nex$ and $\nc$ is found by numerical integration of \Eq{nexPopov}, where an iterative procedure is used to find a self-consistent solution. The difference between the semi-ideal model and the Popov model is shown in \Fig{bimodalBenP}, focusing on the density distribution of the condensed atoms. The density distribution of the thermal atoms as shown in \Fig{bimodal} results in a higher effective potential for the condensed atoms. This effect results in a compression of the density distribution of the condensate compared to the distribution calculated using the semi-ideal model.

\subsection{Determining the chemical potential in the experiment}
In the theoretical models the density profile is determined by the parameters $\mu$ and $T$ in a harmonic confinement given by $V_\text{ext}$. Therefore, it seems evident to analyze the measured density profiles by fitting them to a theoretical model in which only $\mu$ and $T$ are to be determined. However, this approach can be applied only if the size and the height of the measured profiles can be determined absolutely. Since the aforementioned PCI method is expected to do so, we can apply such a fit procedure here.

In the commonly used imaging methods the size or the height of the density profiles is not measured absolutely, and one has to fit the size and height of signal of both the BEC and the thermal cloud separately. In that case, $\mu$ and $T$ are derived from the height and size of the measured density profiles. The experimental conditions favor the use of one of these measures above the others. 

Being able to describe the complete cloud by $\mu$ and $T$, as is done in the theoretical models, is expected to yield a higher accuracy, since different measures for the chemical potential and temperature are combined which depend differently on experimental parameters such as the magnification, detuning and the trap frequencies. Furthermore, it allows for taking the interaction between the condensed and thermal atoms into account. This interaction is ignored in the vast majority of experiments in this field, in contrast to the theoretical description of a condensate at finite temperature found in many textbooks. Our PCI setup is expected to yield an accurate and absolute determination of both the density and the size of the cloud and measuring \insitu prevents the need to account for the expansion of the cloud. Therefore, since all measures are absolute, $\mu$ and $T$ are expected to fully describe the measured density profiles, given the fixed experimental parameters such as the magnification and the trap frequencies.  This allows us to directly compare our measured density profiles to the three mean-field models described in the previous section.

The PCI method not only yields absolute density profiles, it is expected to yield a more accurate determination of the thermal fraction for low temperatures compared to absorption imaging as well. Since the accumulated phase for the dense BEC can be multiple times $\pi$, a significant phase is accumulated in the thermal cloud, without saturating the signal in the BEC. In absorption imaging on the other hand, the optical density of the thermal cloud is strongly reduced in order to reduce the optical density of the BEC to the order of \num{4}. Especially for low temperatures, when only the low density tail of the thermal distribution spatially extends the BEC, it turns out to be difficult to distinguish between both components in the experiment. Various schemes have been used to enhance the contrast between the thermal cloud and the condensate, for example by making a two-pass fit routine, first only on the thermal cloud, followed by the condensate and thermal cloud while fixing the temperature found in the first pass \cite{Gerbier2004}. Another scheme used is spatial separation of both components using Bragg spectroscopy and fitting both parts separately \cite{Gerbier2004}. In the PCI method the contrast between both components is sufficient for all measured temperatures and prevents the need to introduce elaborate schemes to enhance the contrast.

\section{Experimental results}
\subsection{Experimental parameters}
Imaging the atoms is conducted in such a way that the periodicity of the intensity of the PCI technique is used. Since the intensity signal varies periodically as a function of $\phi_\text{atoms}$ (\Eq{phase_phasespot_full}), the intensity signal shows rings in the intensity profile for sufficiently large values of $\phi_\text{atoms}$ (\cf \Fig{phase-phi-theta}). The number of rings depends on $\phi_\text{atoms}$, which scales with the density and the detuning $\delta$ as can be seen from \Eqs{realphase} and \ref{eqn:polphi}. The detuning is a parameter which can be tuned accurately in a wide range (up to \SI{400}{MHz}) using acoustic-optical modulators (AOMs) allowing us to make images with an adjustable number of rings. Due to the ring pattern the imaging lens can be put in position very sensitively allowing us to resolve features down to \SI{4}{\micron}. Furthermore, $\delta$ can be chosen in such a way that the intensity profile of the BEC shows rings and the less dense thermal cloud yields a significant intensity signal as well, even for low temperatures. The resulting effective dynamic range of this method is therefore increased by the periodic dependence of the intensity on the accumulated phase. 

A drawback of this method lies in the fact that the BEC acts as a gradient index lens, the lensing being stronger for smaller $\delta$. The effect of the lensing on the imaging resolution is estimated by a computer simulation in which the Fresnel-Kirchhoff diffraction integral is numerically integrated. In this simulation the finite size of the imaging lenses is explicitly taken into account.  The mutual distances between the BEC, lenses, phase spot and image plane are chosen identical to the values used in the experiment. The intensity distribution at the image plane is used to estimate the effects of the refraction of the BEC on the imaging resolution. If the diffraction is too large, the higher diffraction orders can miss the aperture of the imaging lenses and will degrade the resolution of the image. This simulation yields an upper limit for the allowed phase shift of the BEC and therefore a lower limit of the applied detuning given the condensate density and dimensions. In the experiment we are in the regime in which the aperture of the lenses is not limiting the resolution, although the onset of a slight deformation of the density profile is observed for the highest condensate densities. For the typical parameters, the maximum focal length of the BEC is found to be of the order of \SI{100}{\micron}. Even though the diffraction is not too large for the apertures in the imaging path, the imaging lenses have to be placed in focus within a few tens of micrometers to prevent distortions of the imaged intensity profile, since the imaged object itself acts like a lens.

The phase spot has to be aligned  accurately as well, since all of the plane wave light has to propagate through the phase spot. In the long axis of the cloud the diffraction is weak, and for too large a phase spot the diffracted light propagates through the phase spot as well. In order to make a off between  the two criteria that the phase spot should be large enough that the non-diffracted light is phase-shifted, but small enough that the diffracted light is not, the phase spot has a size of only a few times the waist of the probe beam. As a consequence, the phase spot is easily misaligned. Misalignment of the phase spot leads to reflection of the probe beam on the edges of the phase spot and possibly shifting the phase of the light, which is diffracted by the atoms. As a result of the misalignment we notice a higher, tilted intensity profile, which is easily detected, since we have several maxima in the intensity profile. The common way to apply PCI is by using a large detuning and the intensity signal never reaches the first maximum value. Without a well-defined maximum in the image, (slightly) more intensity due to misalignments may be unnoticed and the method can no longer be used to determine the absolute phase.

As can be seen in \Fig{phase-phi-theta}, the intensity as a function of the phase varies in certain ranges rapidly. Choosing a detuning in such a way that the maximum phase lies in this range, a slight increase in the density leading to a slight increase in the maximum phase leads to a large change in the intensity in the center of the cloud making the method very sensitive to small density changes.

In the experiment imaging the atoms consists of taking a sequence of three images. The first image obtains the intensity profile $I_\text{atoms}(x,z)$ of the probe field and the atoms. The second image contains the intensity profile $I_0(x,z)$ of the probe field in absence of the atoms. This image is generally shot after the magnetic trap has been turned off for two seconds and no atoms are left. The third image is the intensity profile $I_\text{bg}(x,z)$ with neither the probe field nor the atoms and is used as a background image. The normalized intensity profile $I(x,z)$ of the atoms is given by
\begin{equation}
  I(x,z) =  \frac{I_\text{atoms}(x,z)-I_\text{bg}(x,z)}{I_0(x,z)-I_\text{bg}(x,z)},
  \label{eqn:ccdimage}
\end{equation}
\noindent where the subtraction and division is performed on a pixel-by-pixel basis. 
The magnification scheme used causes the probe beam to get blown up to such an extend that the beam profile is no longer smooth on the camera due to interference effects caused by dust and small scratches on the imaging optics. In principle, these imperfections are canceled in the final images, but since the imaging path is not interferometrically stable in between the acquisition of the three images, some distortions remain.

\subsection{Phase contrast images}
\begin{figure*}
\begin{center}
\begin{tabular}{cc}
\includegraphics[width=0.48\textwidth,angle=0]{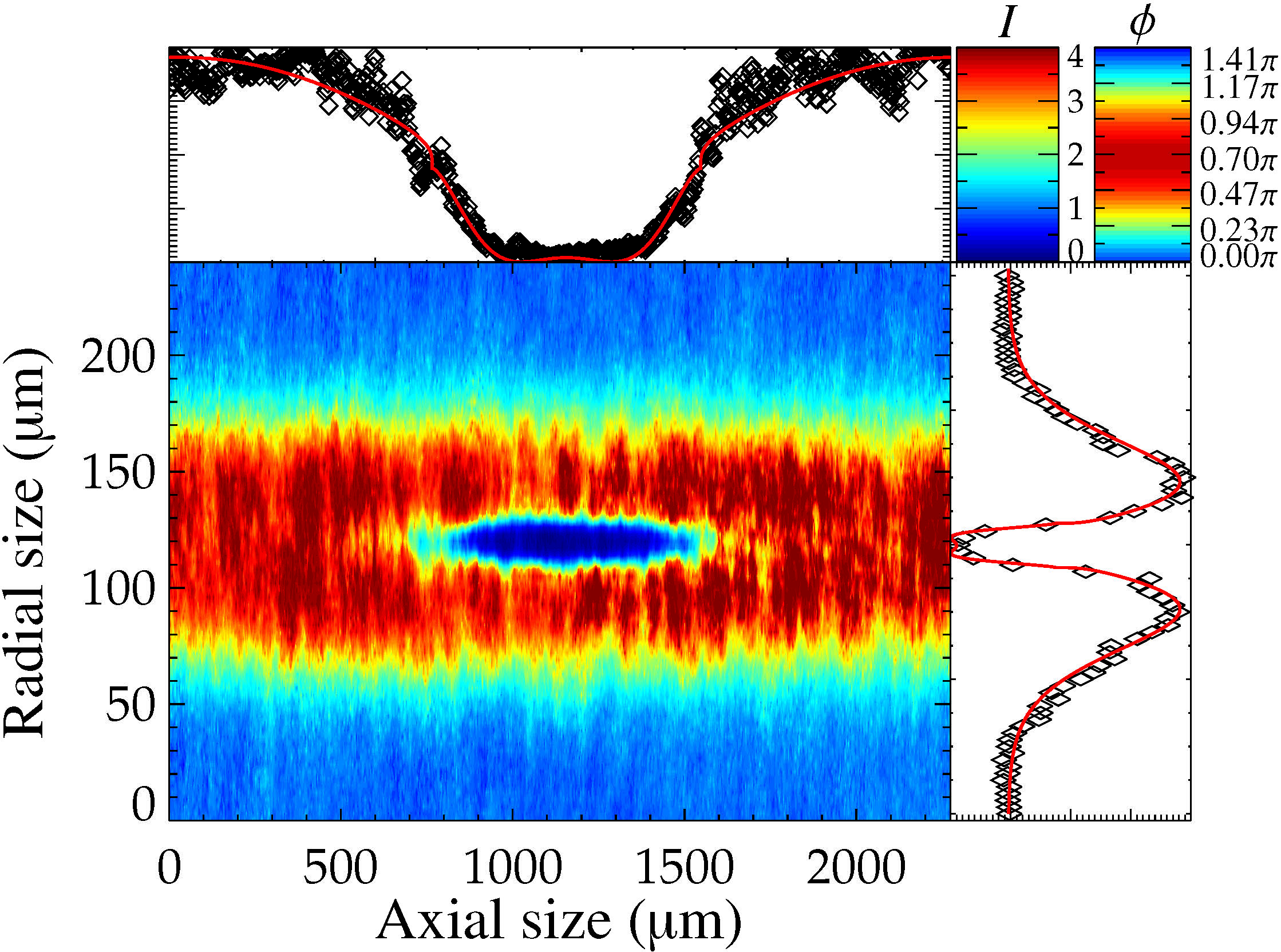} &
\includegraphics[width=0.48\textwidth,angle=0]{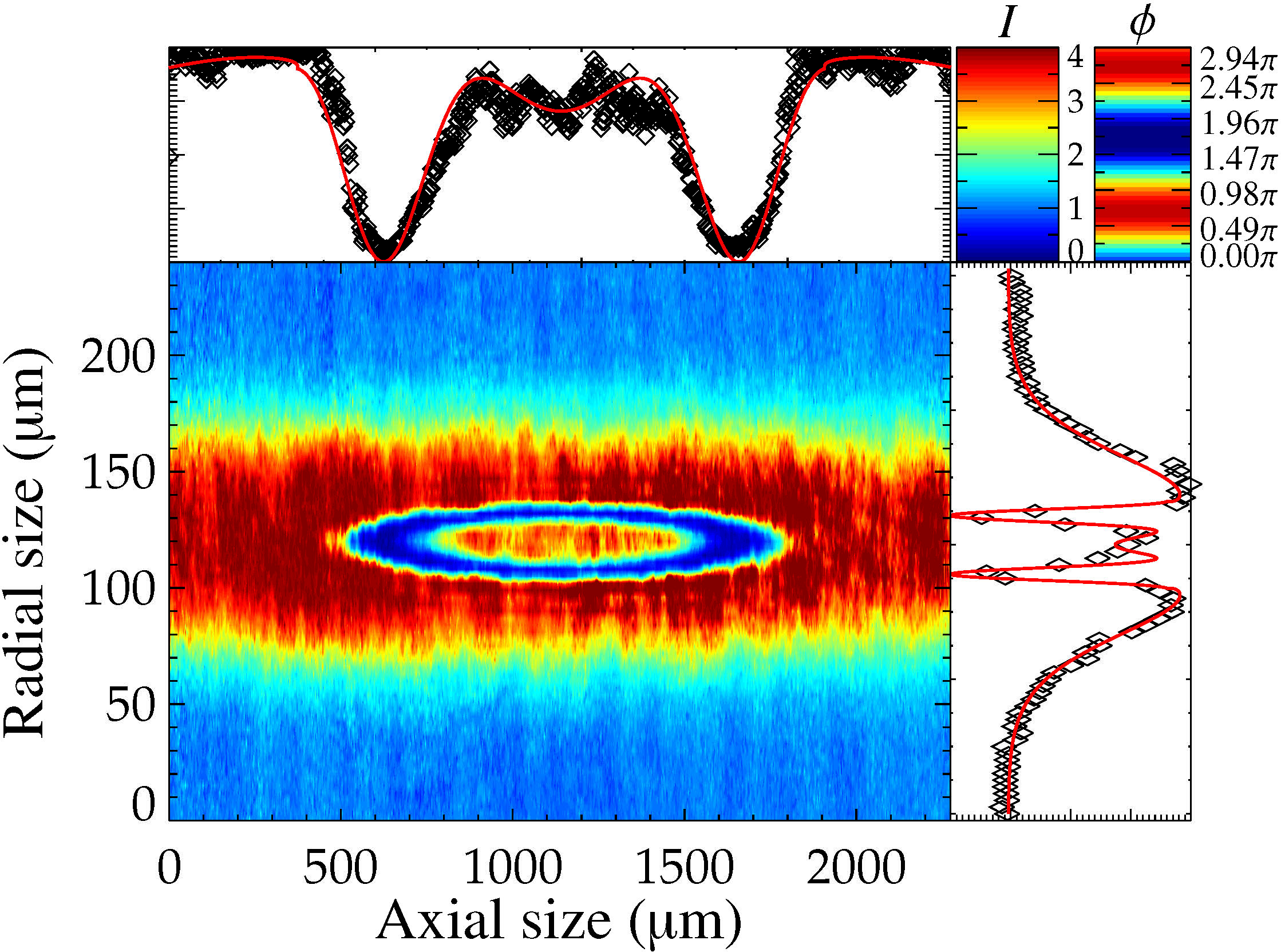} \\
(a) & (b) \\
\includegraphics[width=0.48\textwidth,angle=0]{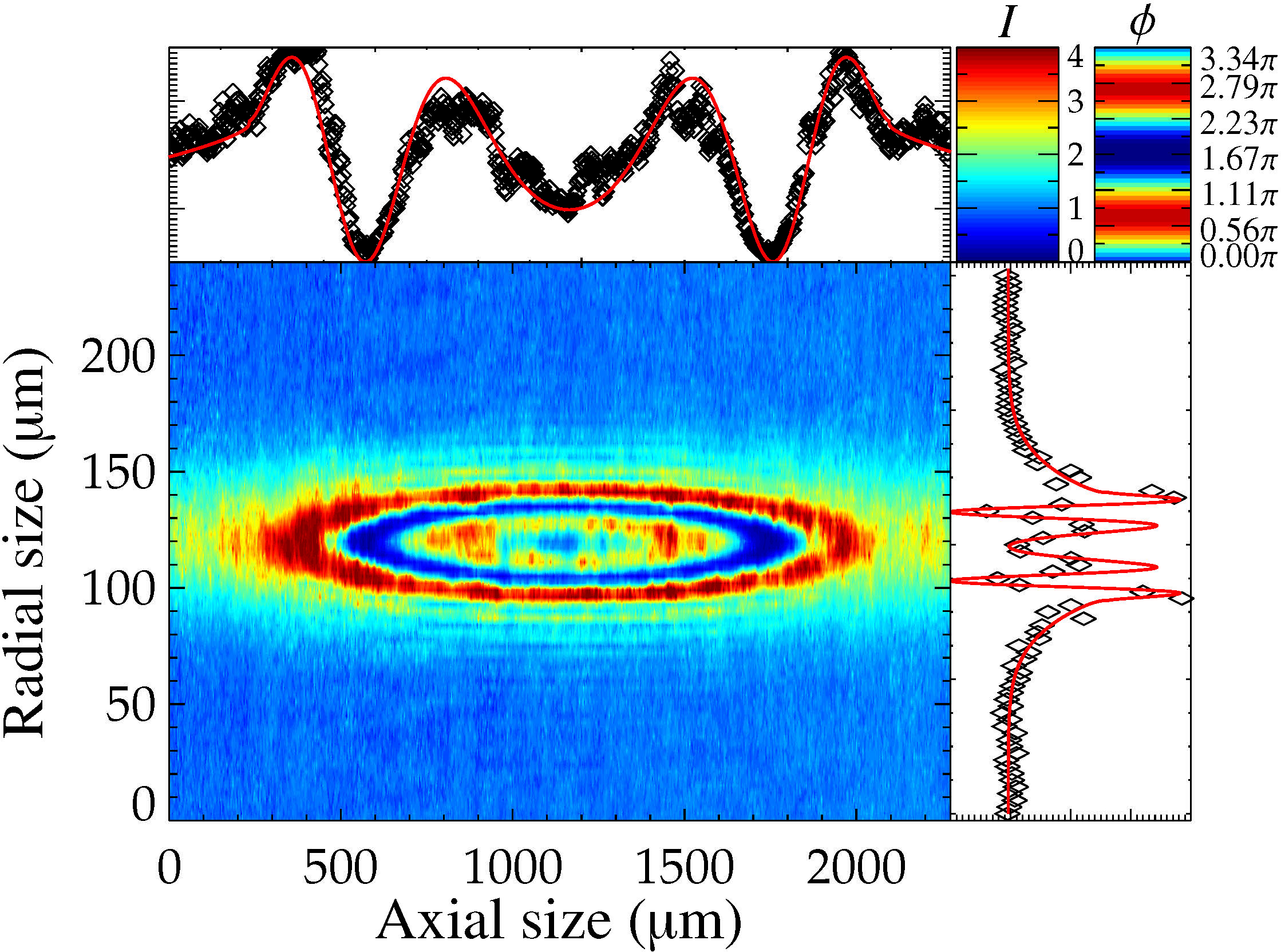} &
\includegraphics[width=0.48\textwidth,angle=0]{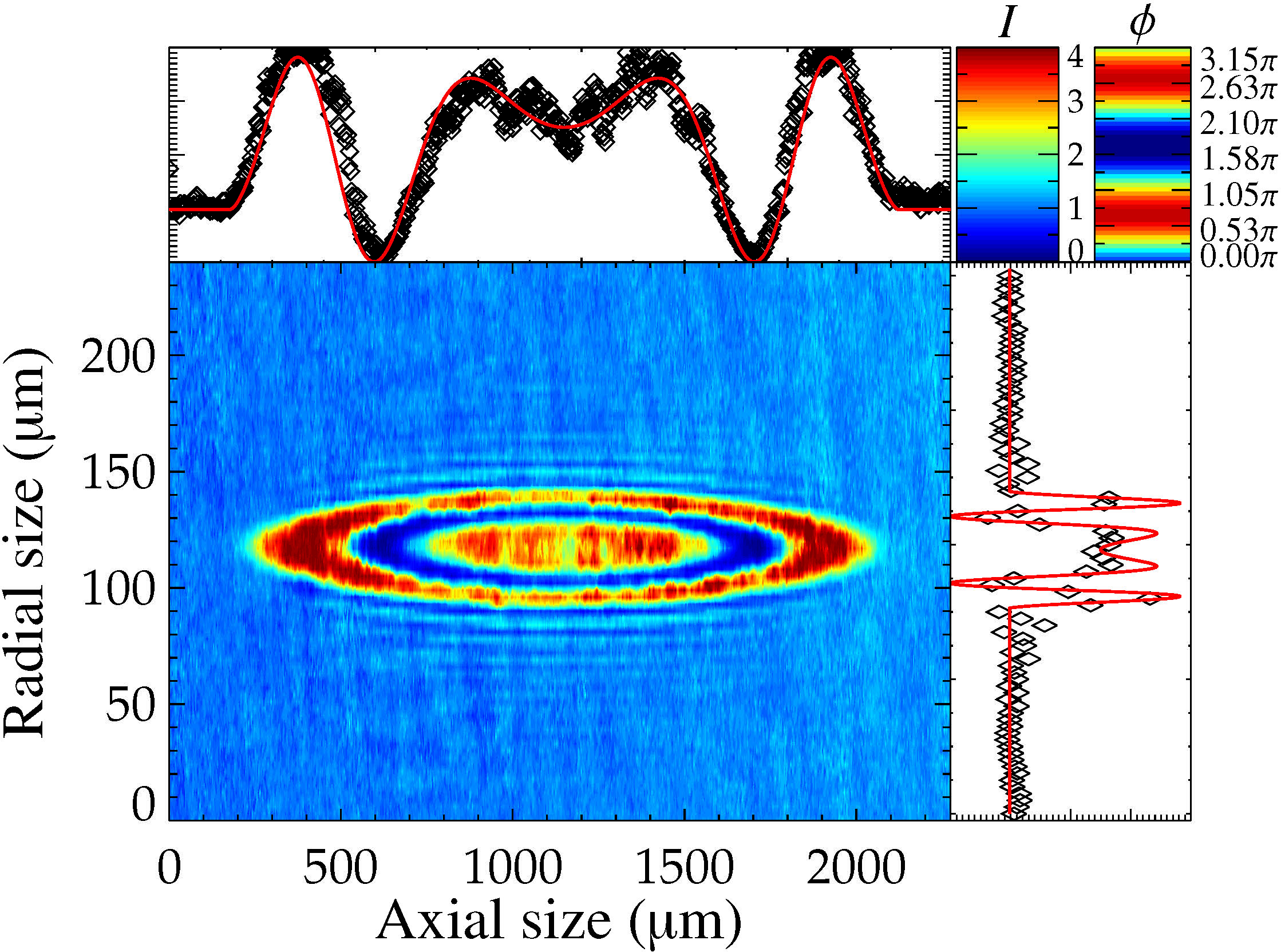} \\
(c) & (d) 
\end{tabular}
\end{center}
\caption{A set of four phase contrast images during the final stage of the evaporation process:
(a) $\mu/h=\SI{2.6}{kHz}$, $T=\SI{1.01}{\microk}$, $N=\SI{7\cdot10^8}{atoms}$, $\Nc/N<\SI{1}{\%}$;
(b) $\mu/h=\SI{4.4}{kHz}$, $T=\SI{0.95}{\microk}$, $N=\SI{6\cdot10^8}{atoms}$, $\Nc/N\approx\SI{15}{\%}$;
(c) $\mu/h=\SI{5.2}{kHz}$, $T=\SI{0.58}{\microk}$, $N=\SI{4\cdot10^8}{atoms}$, $\Nc/N\approx\SI{50}{\%}$;
(d) $\mu/h=\SI{4.8}{kHz}$, $T=\SI{0.25}{\microk}$, $N=\SI{2\cdot10^8}{atoms}$, $\Nc/N>\SI{95}{\%}$.}%
\label{fig:serie21}
\end{figure*}

To demonstrate the PCI method, atoms are cooled below the critical temperature $T_\text{c}$ in a $\SI{95.57}{Hz}\times \SI{95.57}{Hz} \times \SI{2.234}{Hz}$ trap and imaged with a probe beam with an intensity $I_0= \num{6\cdot10^{-2}}\,I_\text{sat}$ for \SI{50}{\microsec}. The probe beam is detuned $\num{28.1}\gamma$ below the $\ket{\Fg=1}\rightarrow \ket{\Fe=1}$ transition and results in typical intensity profile $I(x,z)$ as shown in \Fig{serie21}. The high frequency noise found in these images is caused by the distorted profile of the probe beam. Small fringes close to the BEC are the result of the lenses being slightly out of focus. 

\subsection{Accurate determination of the number of Bose-condensed atoms}
\label{section:problemsabsim}
\subsubsection{Accuracy using absorption imaging}
\label{subsection:problemsabsim}
The usual way to determine the number of condensed atoms is by taking a series of absorption images in time-of-flight. Absorption imaging turns out to become unreliable when the optical density exceeds \num{4} due to the limited dynamic range of the CCD camera \cite{KetterleVarenna1999}. If the cloud of atoms is a BEC, the typical optical density is in the order of \num{500} when the probe is on-resonance. Imaging the atoms off-resonance to take advantage of the reduced photon scattering cross section turns out to be complicated since the cloud behaves as a gradient-index lens in this regime. The optical density can be lowered by turning off the confinement causing the atoms to expand during a certain time-of-flight, until the density is low enough that the optical density is in the order of \num{4}. However, expansion complicates the interpretation of the measured density profiles, since it consists of the convoluted momentum and spatial distribution of the atoms and the expansion of the cloud cannot be described exactly. 

Furthermore, the switching of the magnetic confinement complicates the retrieval of the column density due to the finite time needed to switch the magnetic fields. In the absence of the magnetic field, the quantization axis of the atoms is no longer well defined causing the atoms to align along small residual magnetic fields during time-of-flight. This changes the effective cross section of the atoms for the applied polarized probe light. The magnitude and direction of the residual magnetic fields are expected to be spatially dependent. Since the atoms fall during time-of-flight due to gravity, the effective cross section is also expected to depend on the time-of-flight duration. Furthermore, the imaging lenses have to be repositioned for each time-of-flight duration. A final complication is that the accuracy of detuning of the probe light is limited to \SI{1}{MHz}, corresponding to a reduction of the absorption up to \SI{4}{\%}. All these effects cause uncertainties in the measured number of atoms up to \SI{20}{\%}. 

A technique is proposed based on the nonlinear response of atoms on the applied probe intensity in order to lower the optical density of the cloud without the use of expansion. This allows for quantitative \insitu absorption imaging, although the precise calibration required has to be done after time-of-flight, reducing the accuracy of this method \cite{OptLett.32.3143}. Moreover, the required reduction of the optical density of the condensate results in a reduction of the optical density of the thermal cloud as well, complicating the determination of both density profiles simultaneously. The high intensity needed to reach the strong saturation regime, $I\propto \text{OD}\times I_\text{sat}\approx \SI{8}{W/cm^2}$, with $I_\text{sat}$ the saturation intensity, complicates the implementation of the method as well.
\subsubsection{Accuracy using PCI}
Using the PCI technique circumvents the issues raised above. The detuning is large compared to the uncertainty, $\delta=\SIfout{281}{1}{MHz}$, and since the measurement is conducted \insitu the confinement is not switched off and the quantization axis of the atoms remains well defined. The Zeeman shift caused by the confinement is less than \SI{2}{MHz} and therefore negligible compared to the detuning. 

An image of a BEC taken with PCI technique yields the accumulated phase and the size of the condensate. Analyzing all known systematic and statistical errors show that the largest contribution to the error in the size originates from the magnification of the imaging system. We determined the magnification to be $M=\numfout{3.05}{0.05}$. Since the condensate radius $R$ is proportional to $M$, the number of condensed atoms scales as $\Nc\propto M^5$ (\cf \Eq{TFrmu} and \Eq{nbec}). The trap frequencies are derived from a center-of-mass oscillation measurements yielding the trap frequency with a statistical uncertainty below \num{10^{-3}}. 

The largest contribution to the error in the accumulated phase originates from the lensing effect of the condensate and is estimated to be less than \SI{5}{\%} in the measured phase based on the computer simulation we conducted. This effect does not influence the axial size of the cloud since the density varies slowly in this direction making the lensing negligible. Since the number of atoms scales with the accumulated phase as $N\propto \phi^{5/3}$, lensing is expected to yield an error up to \SI{8}{\%} in the number of atoms for the highest densities.

We analyze measured clouds for similar parameters as the ones shown in \Fig{serie21}, where the condensate fraction is at least \SI{90}{\%} by making a least square fit to the ideal model. We conclude both the determined size and phase of the condensate yield the same chemical potential $\mu$ within \SI{3}{\%} for the typical number of condensed atoms $\Nc \sim \num{2.5\cdot10^8}$. Since the thermal fraction is small, interactions are expected to be of minor importance in this regime.  

We expect that the small discrepancy between the measures for $\mu$ is caused by the lensing effect of the condensate, which slightly alters the density profile. If the imaging lens is aligned incorrectly, more lensing is observed and the discrepancy between the chemical potential based on the phase and sizes increases.

\subsection{Observation of interactions between the thermal cloud and the BEC}
\label{subsection:interactions}
The effect of interactions between the thermal cloud and the BEC becomes important if a significant thermal fraction is present. As pointed out in Ref.~\cite{Giorgini1997}, the thermodynamic behavior of the cloud is fixed by two parameters: the reduced temperature $t=T/T_\text{c}^0$ and the ratio $\eta$ given by 
\begin{equation}
  \eta = \frac{\mu_0^\text{TF}}{\kb T_\text{c}^0} \approx 1.57 \left ( N^{1/6} \frac{a}{\bar{a}} \right )^{2/5}.
  \label{eqn:etaGiorgini}
\end{equation}
Here $T_\text{c}^0= \hbar \bar{\omega} (N/\zeta(3))^{1/3}$ denotes the transition temperature in the absence of interactions and $\bar{a}=\sqrt{\hbar/(m \bar{\omega}})$ is the harmonic oscillator length.

Under our typical conditions $\eta=\num{0.3}$, which is less than its value in other experiments studying the interactions \cite{Gerbier2004,Busch2000}. There it is shown that interactions between the atoms shift the transition temperature $T_\text{c}$ downward. Ref.~\cite{Gerbier2004} measured the effects of the interactions by spatially separating the thermal cloud from the BEC using Bragg spectroscopy to avoid the need to incorporate interactions in the description of the measured density profiles. The separation of both clouds depends sensitively on the applied Bragg pulse, and introduces an extra uncertainty. Furthermore, the accuracy is limited by the absence of an exact theory describing the expansion of the BEC.

Using PCI these interaction can be measured directly and more accurately than in these previous studies, although $\eta$ is smaller. The measurements are conducted as follows. Using evaporative cooling we obtain a cold cloud of atoms at a temperature $T$ below $T_c$ under the same experimental conditions as the measurement shown in \Fig{serie21}. A phase contrast image of this elongated cloud is taken with each cloud cooled to different temperatures. Each image is fitted to all three \twoD models described in \SubSec{threemodels} and yield a temperature and chemical potential. We find the best fit can be made, determined by the smallest sum-of-squares, using the Popov model and the semi-ideal model. The noninteracting model gives inferior results and cannot be used to accurately describe the measured profiles. This has already been noticed in previous work \cite{Gerbier2004,Busch2000}.

In this experiment we can also see the effect of the mean-field potential of the thermal cloud on the condensed atoms although less prominent than the modification of the thermal distribution. This effect is already hinted at in Ref.~\cite{Gerbier2004}. The density distribution of the thermal cloud causes an effectively larger potential for the condensed atoms. The effect of the resulting distribution is analogous to increasing the trap frequency: the cloud gets smaller and denser. Since we can measure both density and size separate and with a high accuracy, we measure the effect of the interactions by comparing the chemical potential based on the axial size and the accumulated phase. The axial size is favored above the radial size for two reasons: lensing effects in this direction are negligible and the size is large compared to the resolution of the imaging setup.

\begin{figure}
\begin{center}
\begin{tabular}{cc}
    \includegraphics[width=0.48\columnwidth]{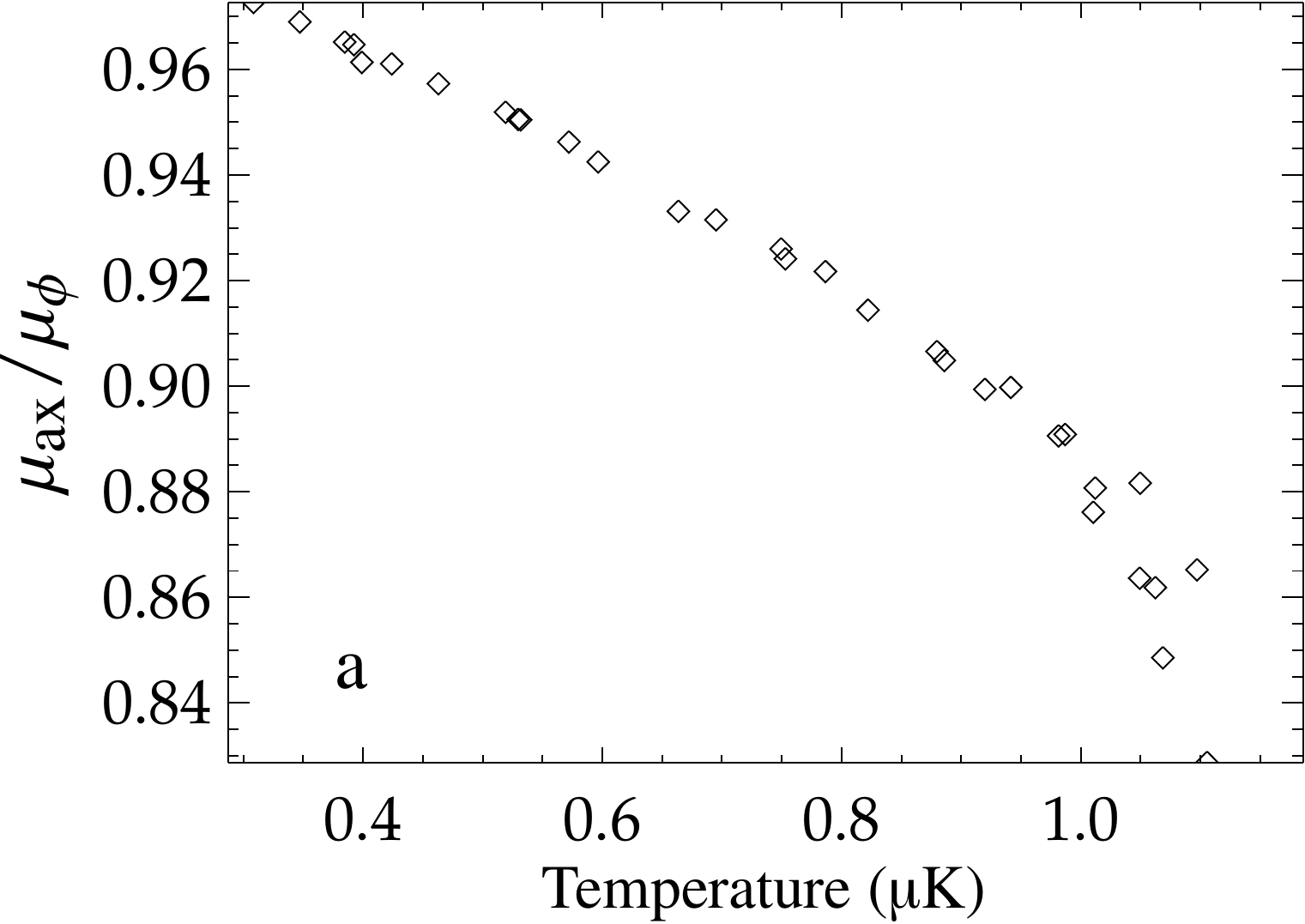} &
    \includegraphics[width=0.48\columnwidth]{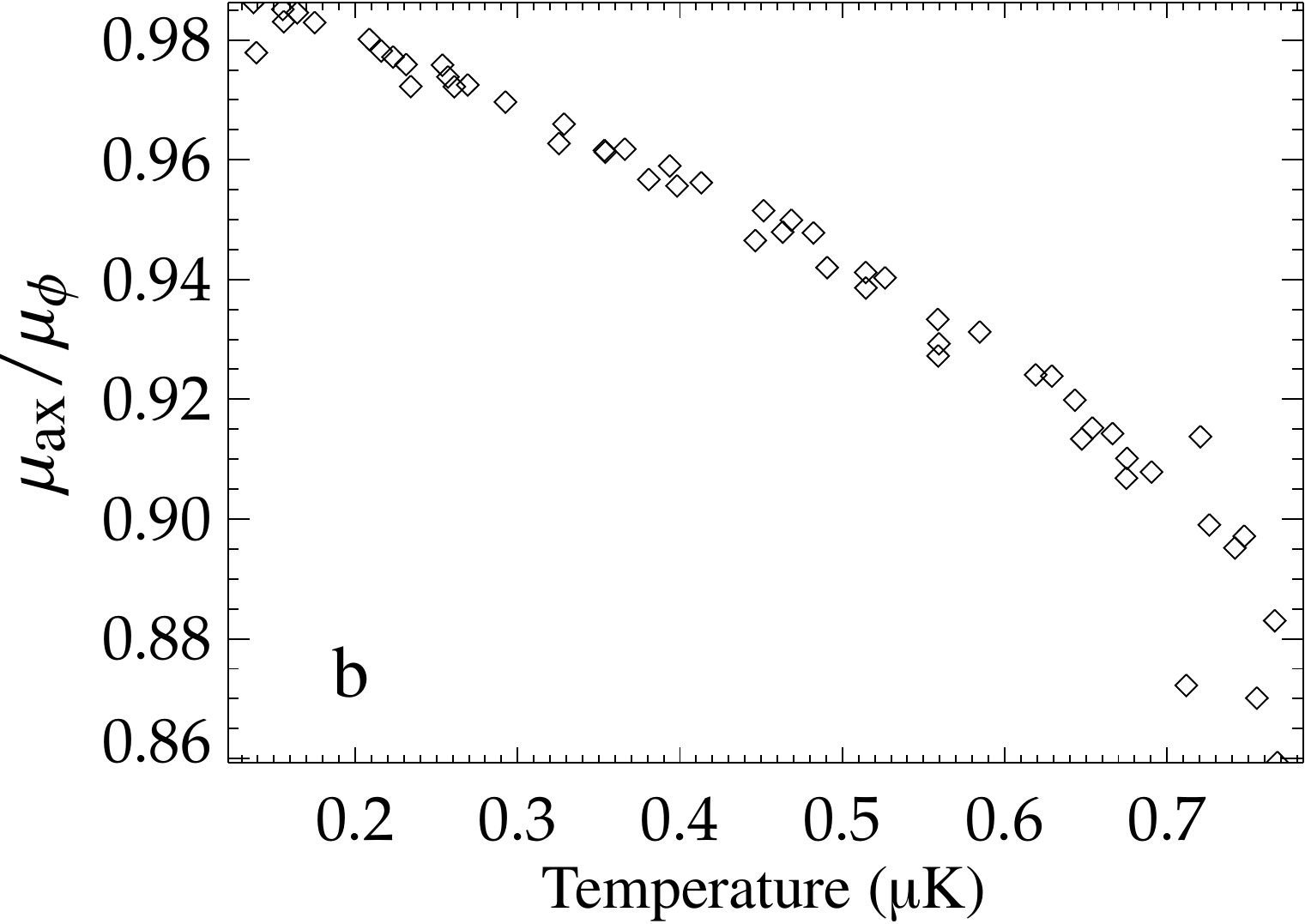} \\
    (a) & (b) 
\end{tabular}
\end{center}
\caption{The ratio $\mu_\text{ax}/\mu_\phi$ of the chemical potential based on the axial size $R_\text{ax}$ and phase $\phi$ as function of the temperature. Figure (a) corresponds to $\eta=\num{0.3}$ and (b) corresponds to $\eta=\num{0.25}$. In the second series $\eta$ is reduced by reducing the number of atoms.}
\label{fig:muellmuax}
\end{figure}

From the fit results we determine the axial size and phase of the condensed part of the cloud and calculate its chemical potential based only on the accumulated phase, yielding $\mu_{\phi}$ or the axial size, yielding $\mu_\text{ax}$. The ratio $\mu_\text{ax}/\mu_{\phi}$ is plotted against the temperature $T$ and shown in \Fig{muellmuax}. We find the chemical potential based on the axial size for temperatures close to $T_\text{c}$ to be approximately \SI{15}{\%} too small compared to the chemical potential based on the accumulated phase. As the temperature decreases, this effect gets smaller until at low temperatures (low thermal densities) we find correspondence between both measures within a few percent, consistent with the results found in the previous section. Since our standard fitting procedure uses $\mu$ and $T$ to fully describe the density profile, we find the sum-of-squares to be \SI{10}{\%} smaller at temperatures where a significant thermal fraction is present (condensate fractions up to \SI{80}{\%}) for the fit using the Popov model compared to a fit using the semi-ideal model. 

Since this shows that the Popov model yields the most consistent results all measurements in the remainder of this paper are analyzed using the Popov model.

\subsection{Accuracy and reproducibility}
Repeated measurements are done in the regime where three-body losses limit the density of condensed atoms. The total number of condensed atoms $\Nc$ is therefore approximately constant and the results are used to determine the sensitivity as well as the reproducibility of the PCI method. The measurement series takes over two hours and during this time the temperature of the cloud changes from $T=\SIfout{0.573}{0.004}{\microk}$ to $T=\SIfout{0.387}{0.004}{\microk}$ due to the apparatus heating up. However, in the density limited regime the number condensed atoms is roughly constant in this temperature range. The measured number of condensed atoms is found to change less than \SI{3}{\%} during the two measurement time on a cloud containing $\Nc=\numfoute{250}{10}{10^8}$ condensed atoms. For measurements shot back-to-back the number of condensed atoms changes approximately \SI{1}{\%} . Since we assume the largest contribution is caused by the difference in the environmental conditions, this result sets the upper limit of \SI{1}{\%} for the shot-to-shot reproducibility of the PCI method using the two parameter fit. 
\begin{figure}
  \begin{center}
    \includegraphics[width=0.75\columnwidth]{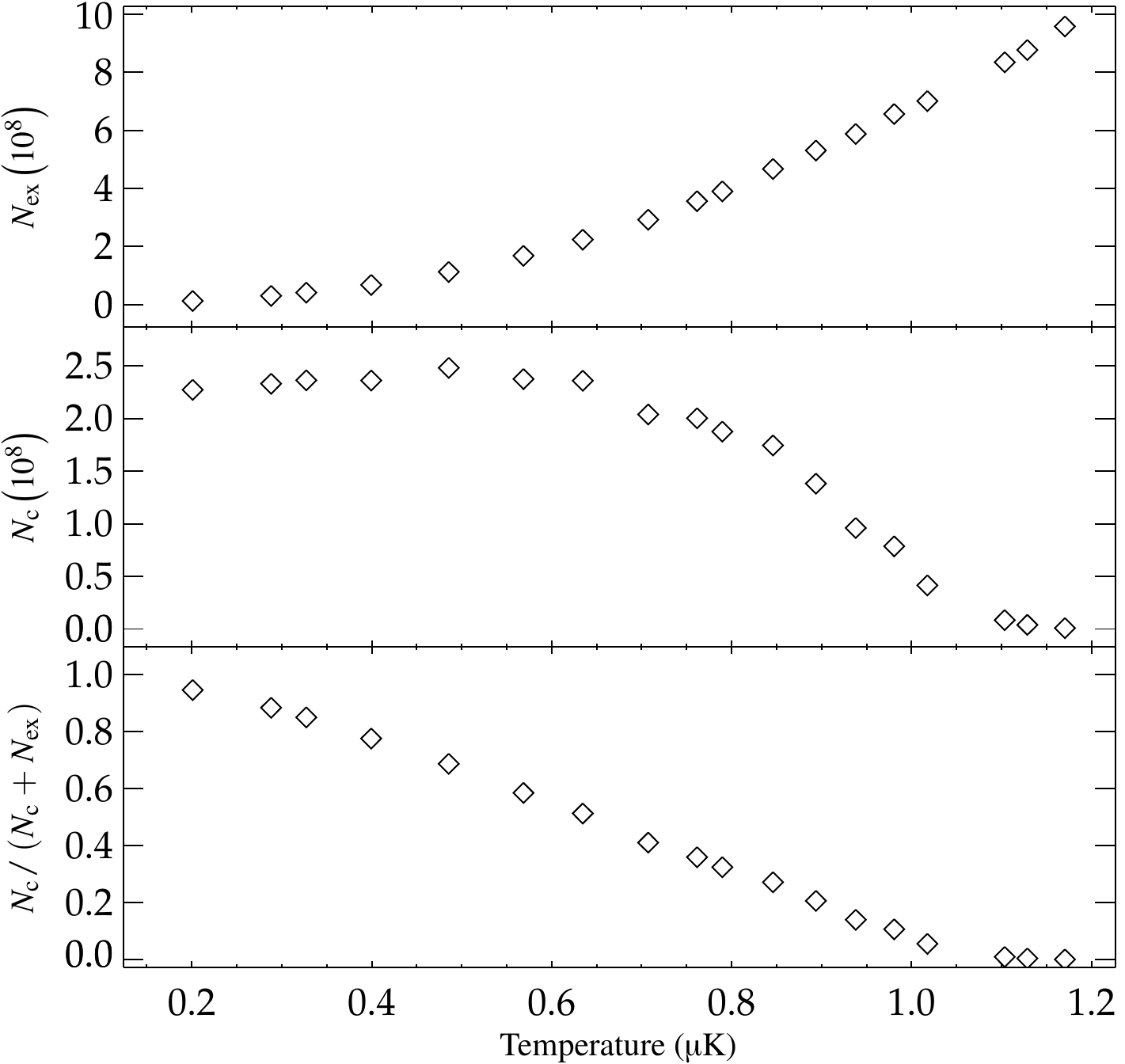}
  \end{center}
  \caption{The number of thermal atoms $\Nex$, condensed atoms $\Nc$ and the condensate fraction as a function of the temperature $T$. Each point is the result of a fit to the Popov model on a single measurement. The statistical uncertainty is smaller than the size of the symbol. }
  \label{fig:condfracpci}
\end{figure}
 
Our fit procedure, in which $\mu$ and $T$ are determined by both phase and size simultaneously using the Popov model, yields statistical errors below one percent, smaller than the estimated systematic error. The uncertainty in the number of condensed atoms, which scales as $\Nc \sim \mu^{5/2}$, is therefore estimated from the discrepancy between $\mu_\text{ax}$ and $\mu_{\phi}$, which is roughly \SI{3}{\%} and yields an uncertainty in the number of condensed atoms of \SI{5}{\%}.

In an experiment we set up to determine the accuracy of the method on a single shot for various temperatures, we produce clouds at various temperatures below \Tc. The resulting number of thermal and condensed atoms, as well as the condensate fraction determined on the fitted $\mu$ and $T$ is shown in \Fig{condfracpci}, where each point corresponds to a single measurement. The fluctuations in these results are only a few percent for both the number of thermal and condensed atoms and this measurement shows the accuracy of the PCI method over the full temperature range.

\subsection{Transition temperature for condensation}
Generating clouds above as well as below the transition temperature allows us to determine the temperature at which condensation sets in. The Popov model yields $\mu$ and $T$ in both regions, where $\mu$ becomes negative for temperatures above $T^0_\text{c}$. Note that our implementation of the Popov model reduces to the Hartree-Fock model for $T>\Tc$ \cite{BECIDG}.

For $T-T^0_\text{c} \ll T^0_\text{c}$ the chemical potential as a function of temperature for a noninteracting Bose gas can be written as \cite{BECIDG}
\begin{equation}
  \mu \approx 3 \frac{\zeta(3)}{\zeta(2)} \kb (T-T^0_\text{c}),
  \label{eqn:pens2.77}
\end{equation}
where $\zeta$ is the Riemann-Zeta function. \Equation{pens2.77} can be used to estimate $T^0_\text{c}$.  Note that the repulsive interactions lower the central density and therefore reduce the transition temperature. Therefore, condensation does not set in at $T^0_\text{c}$, where $\mu=0$, but for $\mu=  2 U_0 \nex$. 

\begin{figure}
  \begin{center}
    \includegraphics[width=0.75\columnwidth]{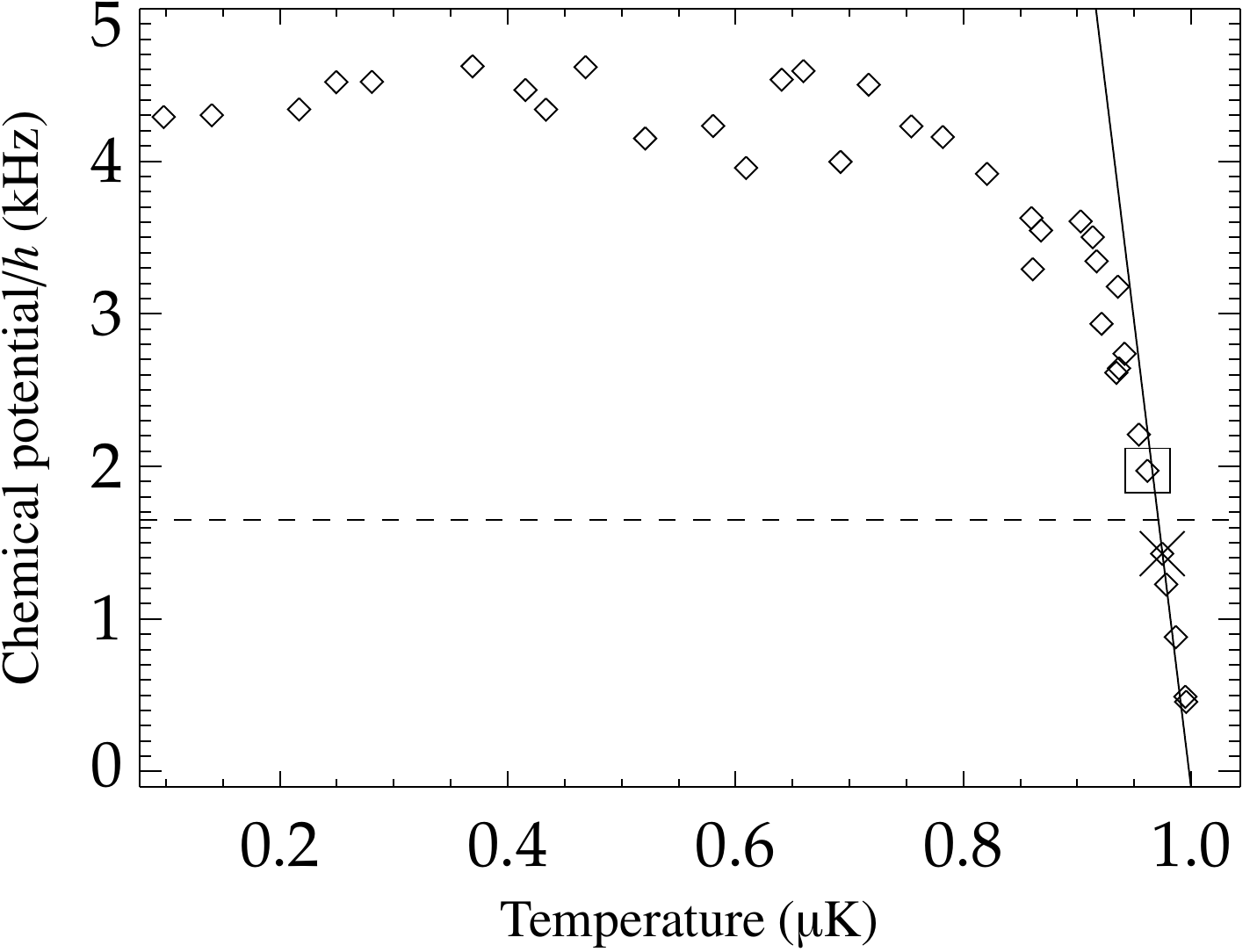}
  \end{center}
  \caption{The chemical potential a cold cloud derived from the fit to the Popov-model as a function of temperature, both above and below $T_\text{c}$. The line is the result of a fit to \Eq{pens2.77} on the seven data points where $\mu/h < \SI{2.5}{kHz}$. The data point accentuated by the box marks the observation of a small BEC. The point accentuated by the cross is the point with the lowest temperature, in which no bimodal distribution is observed. The dashed line indicates $\mu=2 U_0 \nex$, the expected value for condensation to set in, calculated using the average thermal density of the two accentuated points.}
  \label{fig:teecee}
\end{figure}

In the experiment we generate clouds at different temperatures above as well as below the transition temperatures and derive $T$ and $\mu$ from a fit to the Popov model. The results presented in \Fig{teecee} show positive values of $\mu$ for all temperatures, although no BEC is observed in the five points with the highest temperatures. The seven data points with the highest temperature are used to make a fit to \Eq{pens2.77} with $T^0_\text{c}$ being the only free parameter. This yields $T^0_c=\SIfout{0.998}{0.007}{\microk}$. The thermal density is determined by averaging the two measurements which are the closest to the temperature where a condensate is formed. Both $\nex$ and $T^0_c$ are used to find the temperature in \Eq{pens2.77}, where $\mu=2 U_0 \nex=h (\SIfout{1.65}{0.05}{kHz})$. This yields $T_\text{c}=\SIfout{0.972}{0.008}{\microk}$. 

The measured shift due to the interactions is ($T^0_\text{c}-T_\text{c})/T_\text{c} = \numfout{0.027}{0.001}$, which agrees well with the theoretical prediction based on the Hartree-Fock mean-field theory, $\Delta T_\text{c}/T_\text{c} \approx -1.33 a/\bar{a} N^{1/6} \approx \num{0.025}$ \cite{BECIDG}. 

In order to confirm the systematic errors are well estimated, we compare the value of $T_\text{c}$ to the highest temperature where we observe a bimodal distribution in the image, both by fitting the cloud and by examining line profiles through the center of the cloud. The highest temperature where a BEC is observed is determined at $T=\SIfout{0.963}{0.005}{\microk}$, where the lowest temperature without a BEC yields $T=\SIfout{0.975}{0.003}{\microk}$ and the determined value of $T_\text{c}$ indeed lies in between these two temperatures. The consistency of these numbers suggest the systematic errors are well estimated.

\section{Conclusion}
In conclusion, this paper describes the accurate determination of the density distribution of condensates at finite temperatures using PCI. A detailed description is given of the amount of phase a probe beam accumulates when a cloud of atoms is passed. The resulting images show a periodic variation of the intensity as a function of the accumulated phase and the resulting high dynamic range is used to measure both the density distribution of a thermal cloud and a condensate simultaneously at high accuracy. Since all measurements are done \insitu, there is no need to describe the expansion of the cloud. The size of the condensate, also a measure for its density, can therefore be determined accurately as well. The size of the thermal cloud, the measure for the temperature of the cloud, is also determined from the same image. 
Only two relevant parameters, $\mu$ and $T$, are needed to describe the measured clouds given the harmonic confinement as $\mu$ and $T$ determine both the peak density and size of both components of the cloud. This procedure is used to discriminate between three models describing the equilibrium properties of trapped clouds at nonzero temperatures, each model incorporating the interactions between thermal atoms and condensed atoms differently. 

We resolve the effect of the interaction on the density distribution of thermal cloud as well as the smaller effect the interaction has on the density distribution of the condensate. The model accounting for these effects gives consistent results over the complete temperature range.
The imaging scheme is used to determine the number of atoms within one percent from shot-to-shot at our typical number of condensed atoms between \num{2\cdot10^8} and $\num{3\cdot10^8}$.  Systematic errors are estimated by comparing the different measures for the chemical potential and we find the uncertainty in the number of number of condensed atoms to be roughly five percent.
PCI can be used both below and above the transition temperature and we used it to determine $T_\text{c}$ and derived the shift of the transition temperature due to the interactions. The agreement with the theoretical value of this shift indicates the systematic errors are smaller than the uncertainty in the determination of $T$.

\section{Acknowledgments}
This work is supported by the Stichting voor Fundamenteel Onderzoek der Materie ``FOM'' and by the Nederlandse Organisatie voor Wetenschaplijk Onderzoek ``NWO''. We are grateful to Chris R\'{e}tif and AMOLF for the manufacturing of the phase spot.

\end{document}